\documentclass[prb,superscriptaddress,twocolumn]{revtex4-1}
\usepackage{lineno}
\usepackage{epsfig}
\usepackage{graphicx}% Include figure files
\usepackage{dcolumn}% Align table columns on decimal point
\usepackage{bm}% bold math
\usepackage{tabularx}% bold math
\usepackage{hyperref}
\usepackage{tablefootnote}
\usepackage{multirow}
\hypersetup{colorlinks=true, citecolor=blue, filecolor=blue, linkcolor=blue, urlcolor=blue}

\begin{document}
\title{First-principles Investigation of Electrides Derived from Sodalites}

\author{Byungkyun Kang}
\affiliation{Department of Physics and Astronomy, University of Nevada, Las Vegas, NV 89154, USA}

\author{Qiang Zhu}
\email{qiang.zhu@unlv.edu}
\affiliation{Department of Physics and Astronomy, University of Nevada, Las Vegas, NV 89154, USA}

\date{\today}

\begin{abstract}
Recently, the electride materials, with excess anionic electrons confined in their empty space,  have received a growing attention due to their promising applications in catalysis, nonlinear optics and spin-electronics. However, the utilization of electride materials is limited by their thermal instability. Here we introduce an alternative way to achieve the localized anionic electron states via the removal of high symmetric Wyckoff sites of anions from the existing sodalite compounds. Using four halide sodalites as the parental structures, our simulation reveals that the materials after the removal of anionic halide sites exhibit typical electride behaviors that are characterized by the existence of localized electronic states near the Fermi level. Compared to most previously studied electrides, these materials are expected to be more thermally stable due to the complex structural framework and thus promising for practical applications. Among them, Na$_4$(AlSiO$_4$)$_3$ manifests magnetic electronic structure. We demonstrate that this magnetism originates from a highly localized excess electron state surrounded by electronpositive alkaline cations. Our results suggest Na$_4$(AlSiO$_4$)$_3$ could be a promising spintronics component, thus encouraging further experimental study. 
%This results in strong confinement of the excess electrons. 

\end{abstract}

\vskip 300 pt
\maketitle

\section{INTRODUCTION}
Electides represent a unique class of materials in which the excess electrons are trapped inside crystal cavities and serve as the anions \cite{Dye-Science-1990, li2004theoretical, dye2009electrides, hosono2021advances, liu2020electrides}. Localization of electrons provides the early examples of quantum confinement \cite{dye2009electrides}. These trapped electrons are usually loosely bound and float around Fermi level to form unique interstitial energy bands \cite{inoshita2021floating}, giving rise to interesting transport and magnetic behaviors. Therefore, electrides have attracted growing attention for various material applications. While the first crystalline organic electride was made by Dye and coworkers in 1983 \cite{Ellaboudy-1983-JACS}, the use of organic electrides for practical applications was limited by their thermal instability \cite{dye2009electrides}. In 2003, Hosono and coworkers reported the synthesis of the first thermally stable electride based on the inorganic mineral mayenite (12CaO$\cdot$7Al$_2$O$_3$) \cite{Matsuishi-Science-2003}. The resulting electride Ca$_6$Al$_7$O$_{16}$ (C12A7:2e$^-$), with the excess electrons confined in the zero-dimension (0D) cages, exhibit excellent thermal stability and low reactivity with air, opening a new chapter in electride synthesis and utilization. The discovery of C12A7:2e$^-$ has stimulated many new efforts to search for other inorganic electrides. Recently, a number of new electrides with electrons confined in different geometries, including 2D (Ca2N \cite{Lee-Nature-2013}, Y2C \cite{Zhang-CM-Y2C-2014}), 1D (Y$_5$Si$_3$ \cite{Lu-JACS-2016}, Sr$_5$P$_3$ \cite{Wang-JACS-2017}, Mn$_5$Si$_3$ \cite{Zhang-QM-2017},  La$_8$Sr$_2$(SiO$_4$)$_6$ \cite{Zhang-JPCL-2015}), and 0D (YH2 \cite{mizoguchi2016hydride}), have been increasingly investigated in experiment. In the mean time, many new materials with improved functionality were proposed theoretically thanks to the recent advances in computer simulation \cite{Inoshita-PRX-2014, Y2C-PRB-2015, Tada-IC-2014, Ming-JACS-2016, Zhang-PRX-2017, burton2018high, wang2018ternary, Zhu-PRM-2019, Qu-ACSAMI-2019, ZHU20191293, li2021electron}. 
%As a consequence, physical properties represented by the metal–insulator transition and superconductivity emerge in the electride but not for the F-center

\begin{figure*}[ht]
\centering
\includegraphics[width=0.82 \textwidth]{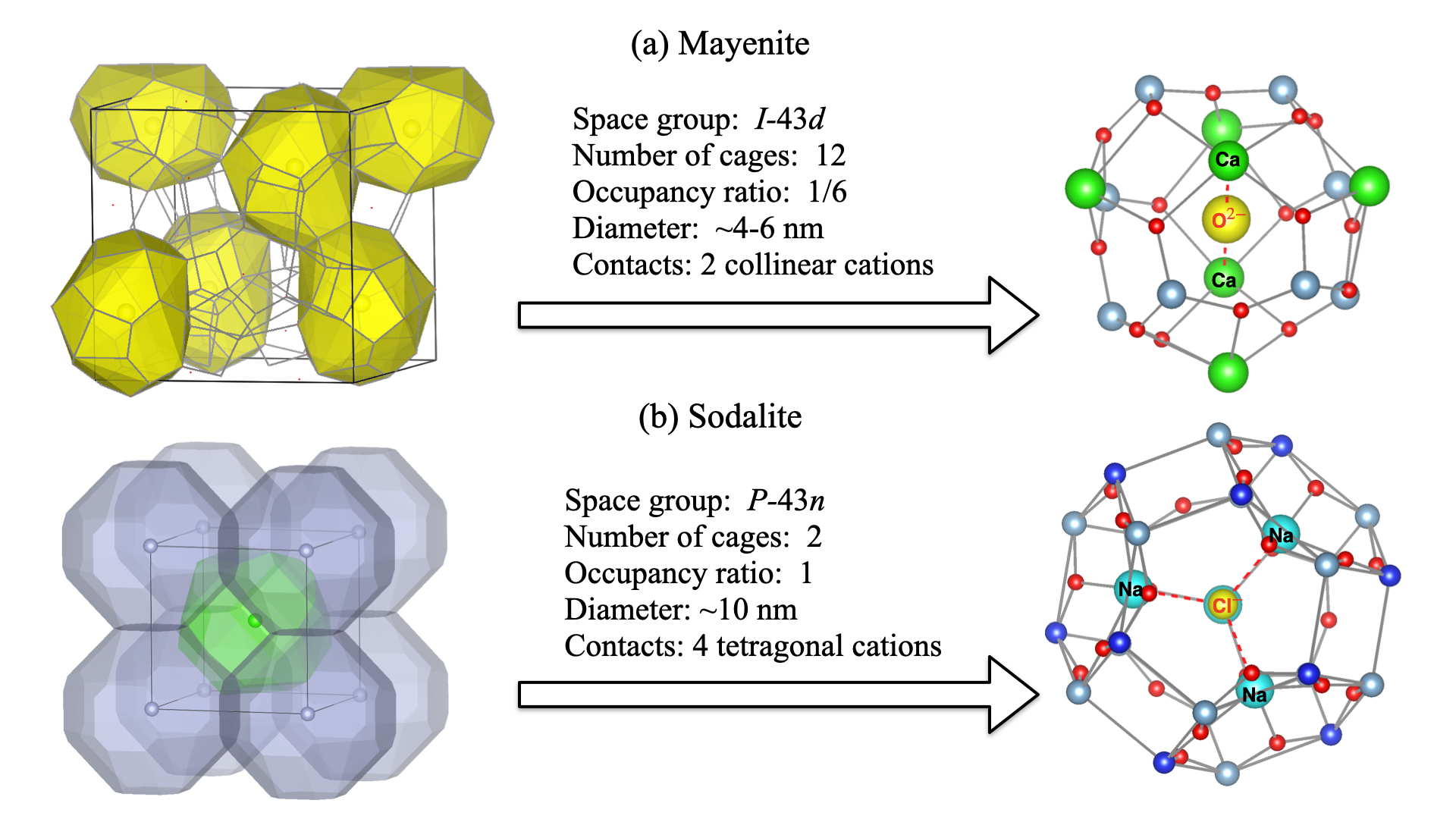}
\caption{\label{Fig1-cages} Two representative porous structures consisting of anions at the center of the cages. (a) mayenite Ca$_12$Al$_14$O$_{33}$ and (b) sodalite Na$_{4}$(AlSiO$_{4}$)$_{3}$Cl. Both structures are featured by the presence of anions (partially) occupying the Wyckoff site around the center of cages. The left panels highlight the overall distribution of cages, while the right panels shows the chemical environment of each representative cage. In each cage, the coordination environment of the center anion is also displayed by the red dash lines. For clarity, only 6 out of 12 cages are shown for the mayenite structure.}
\end{figure*}

%The uniqueness of C12A7 
Despite the growing number of reports on identifying new electride materials, most of them are only stable at low temperatures and very sensitive to air and moisture. To date, C12A7:2e$^-$ remains one of the only room-temperature stable electrides from practical application perspective, with the exception of several recent reports \cite{liu2022preparation, yang-2021-prb, kang2020water}. Therefore, it is necessary to examine the uniqueness of C12A7:2e$^-$ from the viewpoint of structural stability. The original composition of mayenite can be represented as Ca$_{12}$Al$_{14}$O$_{33}$, which is an air-stable neutral material at ambient temperature and atmosphere with a melting point of 1415 $^\circ$C \cite{hosono2021advances}. Its crystal structure has a cubic body centered unit cell, consisting of four positively charged [Ca$_{6}$Al$_{7}$O$_{16}$]$^+$ as the main framework with the following Wyckoff sites: Ca (24d), Al1 (16c), Al2 (12a), O1 (48e), O2 (16c) \cite{boysen2007structure}. As shown in Fig. \ref{Fig1-cages}a, this framework constitutes twelve cages with a free inner space of 4-6 $\textrm{\AA}$ in diameter in the unit cell. In addition, there exist another type of oxygen atoms (O3) that are statistically distributed inside centers of the twelve cages, which can be regarded as the partial occupancy of the 12b Wyckoff site with a ratio of $\frac{1}{6}$. Inside the cage, each O3 site is collinearly coordinated to two Ca atoms with a distance of 2.77 $\textrm{\AA}$, which is a bit higher than 2.40 $\textrm{\AA}$ in the rocksalt CaO compound. The resulting two oxygen may be regarded as the counteranion (O$^{2-}$) to the giant framework cation. Due to the long Ca-O distance, it is expected that O$^{2-}$ anions are loosely bound to the cages \cite{Matsuishi-Science-2003}. When the material is thermally treated in a reductive atmosphere, part of the loosely bonded O$^{2-}$ ions may escape as gaseous O$_2$ from the lattice and leave electrons in the cages, considered as the electron captured oxygen vacancy, or F-center. With the proper reductive condition, all of the loosely bonded O$^{2-}$ anions can be removed from the lattice. In this case, the relatively free electrons play the role of anions and result in a high conductivity. In the meantime, the unpaired feature of these free electrons also generate nontrivial magnetic \cite{Matsuishi-Science-2003} and superconducting properties \cite{Miyakawa-JACS-2007}. Thanks to its excellent stability, C12A7:2e$^-$ has been used for ammonia synthesis \cite{Kitano-NChem-2012, Kuganathan-JACS-2014, Hayashi-JACS-2014} and as an electron-injection barrier material \cite{Hosono-PNAS-2017}.

%What to do with our new 
Inspired by the synthesis of C12A7:2e$^-$, we proceed to check the feasibility of designing new electrides from the existing minerals sharing structural features similar to mayenite that allow the localization of anionic electrons. Clearly, the most important characteristic of a good precursor is the presence of anions occupying the high-symmetry Wyckoff site. In the entire materials database, there exist many compounds meeting such a requirement. In this work, we focus on a common type of sodalite minerals consisting of the halide anions at the high symmetry Wyckoff sites (see Fig. \ref{Fig1-cages}b). Similar to mayenite, the sodalite has a high melting point of 1079 $^\circ$C \cite{antao2002thermal}. Its crystal structure also contains the porous framework with two cages centering around the corner and body center. The halide anions ($X^-$) occupy the cage centers at the 2a Wyckoff site, exhibiting a tetragonal coordination with the neighboring alkaline metals $M^+$ (e.g., Na, K). The resulting $M$-$X$ distance is about the same bond length as that in the typical ionic compound. For instance, the Na-Cl contact in sodalite is 2.76 $\textrm{\AA}$, which is close to 2.79 $\textrm{\AA}$ in the rocksalt NaCl compound. Due to the large dimeter size, it is possible to form many diffusion pathway if the $X$ anion is removed from the $M_4$ tetrahedron to realize the electride state. Using these materials as the precursor, we immediately found that the so called black sodalite Na$_{4}$(AlSiO$_{4}$)$_{3}$ has been found in our recent screening work \cite{ZHU20191293}. In this work, we checked several thermal-stable sodalite compounds and derived the likely crystalline electride structures via rigorous structural analysis. Based on these structures, we further performed a thorough computational investigation on their electronic and magnetic properties based on first-principles calculations. Our results suggest that the electrides derived from sodalite are likely to achieve similar properties as found in C12A7:2e$^-$. We hope this work will encourage more investigations of porous zeolites as the precursor to synthesize more thermal stable electrides in the future.
%Fig. 1, the crystal structures of CAO & Sodalite

\section{Computational Methods}
In this work, we selected four materials for detailed study, including Na$_{4}$(AlSiO$_{4}$)$_{3}$Cl, Na$_{4}$(AlSiO$_{4}$)$_{3}$Br, Na$_{4}$(AlGeO$_{4}$)$_{3}$I and Na$_{4}$(GaSiO$_{4}$)$_{3}$I. For each of them, we first created the new structures by removing the halide anions in the unit cell. The structures are subsequently relaxed to obtain the optimum cell parameters. The structural stability was then checked by phonon calculation. Whenever the imaginary frequencies were found, we followed the largest soft mode to obtain a new structure with a subgroup symmetry\cite{kang_prb2014}. 

\begin{figure*}[htbp]
\centering
\includegraphics[width=0.95 \textwidth]{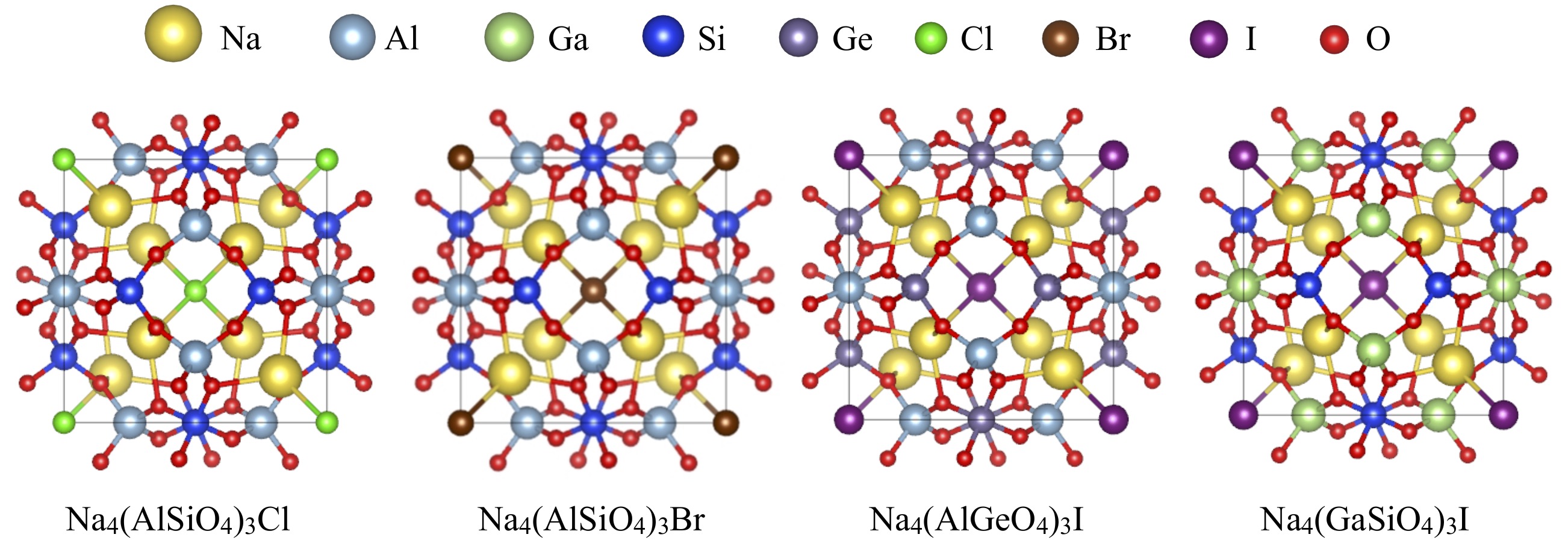}
\caption{\label{Fig_str} The optimized crystal structures of four sodalite materials.}
\end{figure*}

We used the projector augmented wave (PAW) method with the plane-wave code VASP \cite{PAW-PRB-1994, Vasp-PRB-1996, kresse1999ultrasoft} to carry out all calculations with the framework of density functional theory (DFT). The Generalized Gradient Approximation (GGA) with Perdew, Burke and Ernzerhof (PBE) functional \cite{PBE-PRL-1996} was adopted.
We have used a 3$\times$3$\times$3 $\Gamma$-centered  k-point grid for geometric relaxation of unit cell, which contain two formula units. The cutoff energy for all calculations is 520 eV, and forces are converged within 0.03 eV/$\textrm{\AA}$. 
Using the optimized structure, we constructed a 2$\times$2$\times$2 supercell (368 atoms) for the F-center defect (with the halide anion vacancy) calculation, in which a single $\Gamma$ point was used for sampling the Brillouin zone.
Phonon bands were calculated using the force constants, which were calculated by the finite displacement methods using the Phonopy code \cite{Togo-PRB-2008}.

\section{Results and Discussions}
\subsection{Pristine Sodalites}

Sodalite is a common structural type for many nature-occurring minerals.  
The typical crystal structure possesses a cubic symmetry in space group $P\bar{4}3n$ (218). As shown in Fig.~\ref{Fig_str}, each unit cell contain two formula units. 
According to X-ray diffraction analysis \cite{linux_zkri1930,mcmullan_actacryst1996,ishmael_am2004}, the mineral Na$_{4}$(AlSiO$_{4}$)$_{3}$Cl is characterized by a $\beta$-cage made of four-membered (Al, Si)O$_{4}$ rings on the (100) plane linked together in six-membered rings parallel to (100) plane. This results in a highly symmetric framework with just five atoms in the asymmetric unit. Other materials, including Na$_{4}$(AlSiO$_{4}$)$_{3}$Br, Na$_{4}$(AlGeO$_{4}$)$_{3}$I and Na$_{4}$(GaSiO$_{4}$)$_{3}$I, show the same packing behavior except a slight difference in terms of lattice parameters.
%The negative charge induced by the AlSi substitutions is balanced by two tetrahedral clusters.
%As determined by different structural refinements over the past century [4–7], this mineral is characterized by a $\beta$-cage that is made of four-membered (Al,Si)O4 rings on the (100) plane linked together in six-membered rings along the [111] direction, leading to the (AlSiO4)66– framework. The negative charge (–6) induced by the Al3+/Si4+ substitutions is balanced by two tetrahedral (Na4Cl)3+ clusters. The structure is highly symmetric, with just five atoms in the asymmetric unit, i.e., Cl (Wyckoff site 2a) in (0,0,0), Al (6d) in (1/4,0,1/2), Si (6c) in (1/4,1/2,0), Na (8e) in (x,x,x) and O (24i) in (x,y,z).
%The four compounds presented are relaxed in the sodalite framework. 

\begin{table}[ht]
\caption{The comparison of cell parameters and band gaps for sodalite materials between simulation and experimental values.}\label{tab_pro}
\vspace{2mm}
\begin{tabular}{lcccc}
\hline\hline
\multirow{2}{*}{System}     &\multicolumn{2}{c}{~~Cell Parameters (\AA)~~} & \multicolumn{2}{c}{Band Gap (eV)} \\
                            & ~~~~DFT~~   & Expt.                     & ~DFT~             & ~Expt.~          \\\hline
Na$_{4}$(AlSiO$_{4}$)$_{3}$Cl & 8.973  & 8.876 \cite{niels_zeolites1991}  & 4.69 & 6.1 \cite{doorn_jes1972}  \\
Na$_{4}$(AlSiO$_{4}$)$_{3}$Br & 9.031  & 8.932 \cite{niels_zeolites1991}  & 4.61 & 5.9 \cite{doorn_jes1972}  \\
Na$_{4}$(AlGeO$_{4}$)$_{3}$I  & 9.287  & 9.174 \cite{geoffrey_pzmm1997}   & 3.39 &                 \\
Na$_{4}$(GaSiO$_{4}$)$_{3}$I  & 9.198  & 9.074 \cite{geoffrey_pzmm1997}   & 3.59 &   \\
\hline\hline
\end{tabular}
\end{table}

Table~\ref{tab_pro} summarizes the calculated lattice parameters, which agree well with the experimental reports. Using the dispersion corrected calculations with the DFT-D3 method \cite{stefan_jcp2010}, we found that the relaxed lattice parameter of Na$_{4}$(AlSiO$_{4}$)$_{3}$Cl is 8.897 $\textrm{\AA}$, which is approximately 0.1 $\textrm{\AA}$ smaller than the value with the standard PBE functional. This indicates a minor effect of dispersion correction in the sodalites. Our results are close to other DFT results in which the calculated lattice parameters are 8.721\cite{amir_jpcc2018}, 8.91\cite{kendall_jcp1998}, and 8.78 $\textrm{\AA}$\cite{ulian_mdpi2022}.

\begin{figure*}[htbp]
\centering
\includegraphics[width=0.95 \textwidth]{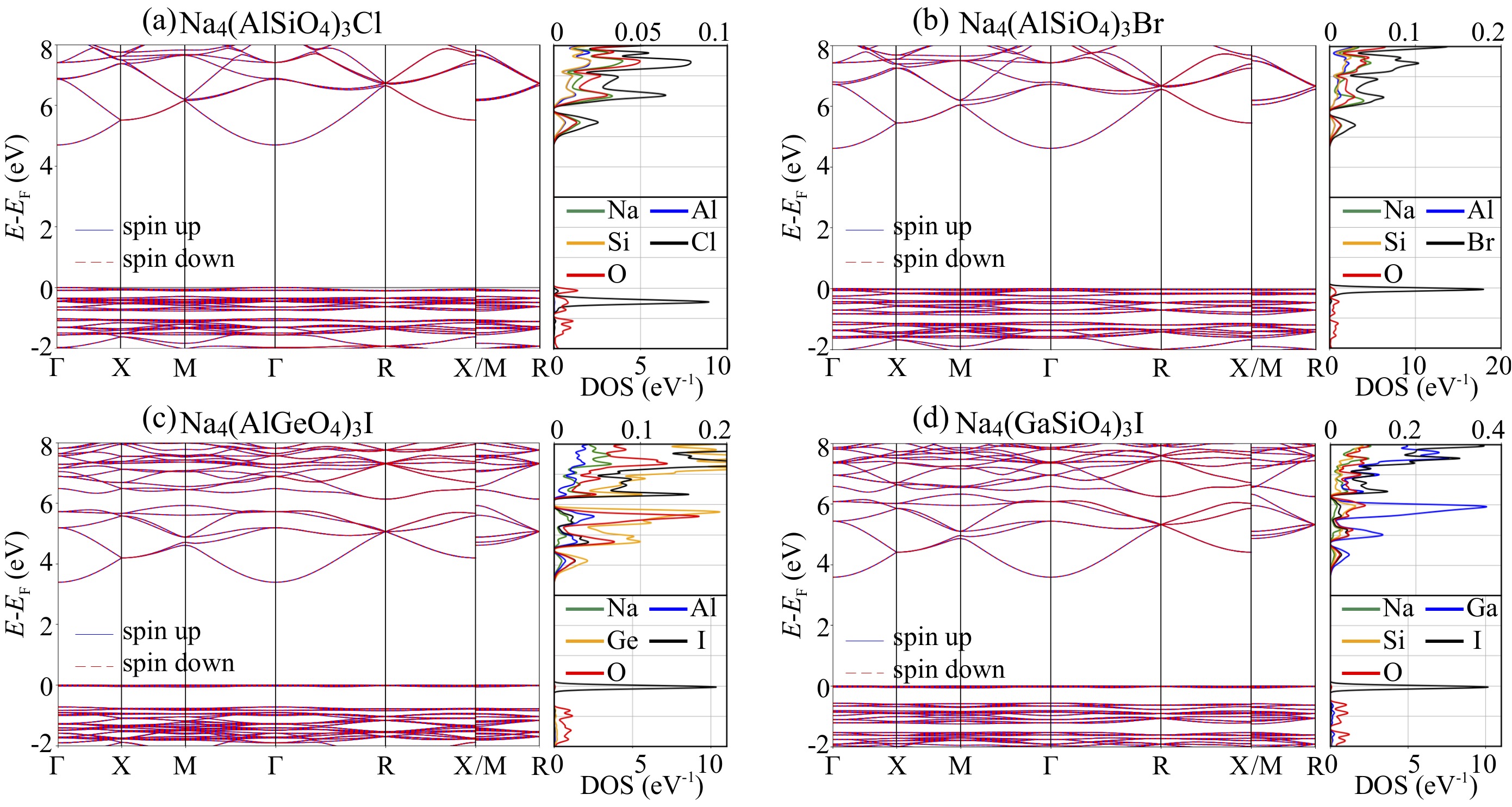}
\caption{\label{Fig_perfect} The electronic band structures and atomic projected density of states of (a) Na$_{4}$(AlSiO$_{4}$)$_{3}$Cl, (b) Na$_{4}$(AlSiO$_{4}$)$_{3}$Br, (c) Na$_{4}$(AlGeO$_{4}$)$_{3}$I, and (d) Na$_{4}$(GaSiO$_{4}$)$_{3}$I. In each band structure, blue solid lines represent spin up bands and red dashed lines represent spin down bands. In the plots of density of states, two different scales of density of states are used: one below 3 eV and one above 3 eV.}
\end{figure*}

Fig.~\ref{Fig_perfect} shows the band structures and density of states (DOS) of all four sodalite materials. Clearly, none of their electronic structures show spin polarization in their band structures.
This means that the electron energy levels are not affected by the spin of the electron.
In addition, all four compounds exhibit direct band gaps at the $\Gamma$ point.
The calculated band gaps are also summarized in Table~\ref{tab_pro}.
For Na$_{4}$(AlSiO$_{4}$)$_{3}$Cl, we obtained a gap of 4.69 eV, similar to other DFT studies (e.g., 4.39 eV \cite{lijun_jm2016, francesco_jpc1995} and 5.0 eV \cite{nilo_ssc2010}). All of these values are smaller than the experimentally reported 6.1 eV \cite{doorn_jes1972}. This is expected since DFT systematically underestimates the energy gap. Similar trends were also observed in other compounds. %However, it captured the insulator nature of these compounds.

From the DOS analysis, the computed valence bands near Fermi level in all four sodalite structures are very flat valence bands, with the contribution of O-$p$ orbitals. On the contrary, their conduction bands are rather dispersive, which are very similar to that of mayenite \cite{hosono2013exploring}. For Na$_{4}$(AlSiO$_{4}$)$_{3}$Cl, its valence band maximum (VBM) consists of large O-$p$ and small Cl-$p$ orbitals, while Na$_{4}$(AlSiO$_{4}$)$_{3}$Br has a VBM comprised of large Br and small O orbitals. This is consistent with the fact that the Br is more electronegative than Cl. For Na$_{4}$(AlGeO$_{4}$)$_{3}$I and Na$_{4}$(GaSiO$_{4}$)$_{3}$I, their highest valance bands are mainly comprised of I orbitals and are separated from the other valence bands. It can be explained by the lowest electronegativity of iodine among the halides (I, Br, Cl), which leads to less hybridization with O-$p$ and a higher energy level.

\subsection{The Structures of Crystalline Electrides}

\begin{figure*}[ht]
\centering
\includegraphics[width=0.99 \textwidth]{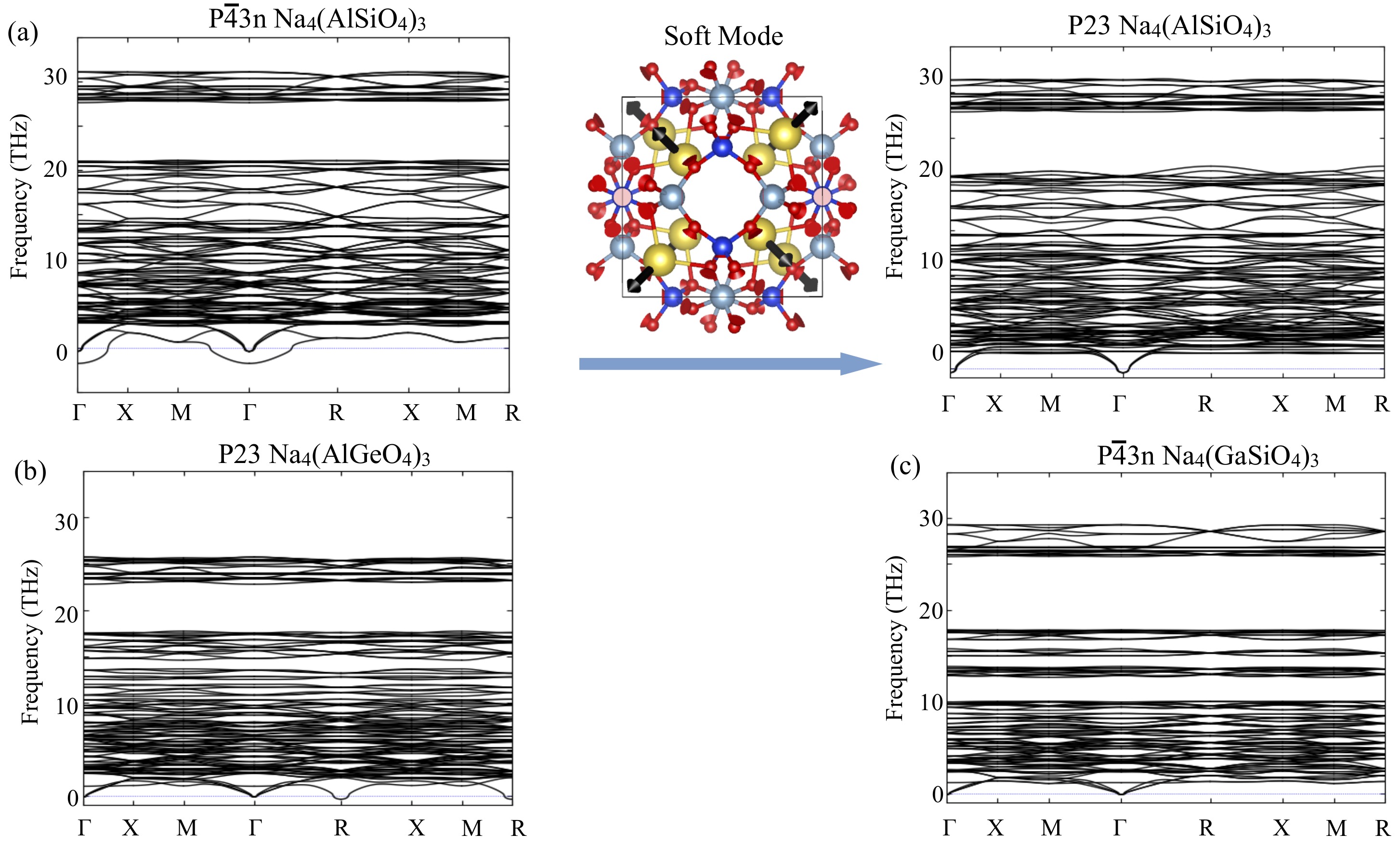}
\caption{\label{Fig_phonon} The computed phonon spectrum for three different electrides (a) Na$_{4}$(AlSiO$_{4}$)$_{3}$, (b) Na$_{4}$(AlGeO$_{4}$)$_{3}$, and (c) Na$_{4}$(GaSiO$_{4}$)$_{3}$.
In (a), the left panel shows phonon dispersion curves of relaxed Na$_{4}$(AlSiO$_{4}$)$_{3}$, which has a soft mode shown in the middle panel. The right panel shows phonon dispersion curves of relaxed Na$_{4}$(AlSiO$_{4}$)$_{3}$ after applying atomic displacements according to the depicted soft mode.}
\end{figure*}

% From F-cener to crystalline electride.
Prior to investigation of electride, an electron trapped at the site of an anion vacancy in ionic crystals is known as an F-center in solid state physics. F-center is a point defect and its concentration is so small that interaction between them is negligible. On the other hand, electrides may be regarded as materials having a stoichiometric F-center, and the electron concentration in the electride is high enough to allow them to interact with each other \cite{hosono2021advances}. In the past, most works attempted to derive the crystal structure by directly removing the anions from the parent crystal \cite{Srdanov-PRL-1998, madsen2001electronic}. However, such treatment failed to take into account the likely structural changes due to complete removal of anions (instead of low-concentration point defects).
Therefore, it is important to examine the resulting crystalline electride in a more rigorous manner.

\begin{table}[htbp]
\caption{The simulated crystallographic data for three crystalline electrides.}\label{tab_xtal}
\vspace{3mm}
\begin{tabular}{ccccc}
\hline\hline
\multicolumn{5}{c}{~~~~Na$_{4}$(AlSiO$_{4}$)$_{3}$, ~~~~Space group $P$23,~~~~ a = 8.98 \AA~~~} \\
\multicolumn{5}{c}{Atomic coordinates} \\\hline
~~~~~Na1~~~~~  & ~~~~4e~~~~ &~~~~0.72~~~~ &~~~~0.72~~~~ &~~~~0.72~~~~ \\
Na2            & 4e         & 0.17        & 0.17        & 0.17        \\
Al             & 6g         & 0.75        & 0.00        & 0.50        \\
Si             & 6h         & 0.25        & 0.50        & 0.00        \\
O1             & 12j        & 0.36        & 0.05        & 0.65        \\
O2             & 12j        & 0.86        & 0.85        & 0.45        \\\hline
\multicolumn{5}{c}{~~~~Na$_{4}$(AlGeO$_{4}$)$_{3}$, ~~~~Space group $P$23,~~~~ a = 9.28 \AA}   \\
\multicolumn{5}{c}{Atomic coordinates} \\\hline
Na1            & 4e         & 0.18        & 0.18        & 0.18        \\
Na2            & 4e         & 0.70        & 0.70        & 0.70        \\
Al             & 6g         & 0.25        & 0.00        & 0.50        \\
Ge             & 6h         & 0.75        & 0.50        & 0.00        \\
O1             & 12j        & 0.65        & 0.93        & 0.65        \\
O2             & 12j        & 0.85        & 0.85        & 0.43        \\\hline
\multicolumn{5}{c}{~~~~Na$_{4}$(GaSiO$_{4}$)$_{3}$, ~~~~Space group $P\bar{4}3n$,~~~~ a = 9.09 \AA} \\
\multicolumn{5}{c}{Atomic coordinates} \\\hline
~~~~~Na~~~~~   & 8e         & 0.18        & 0.18        & 0.18        \\
Ga             & 6d         & 0.25        & 0.00        & 0.50        \\
Si             & 6c         & 0.25        & 0.50        & 0.00        \\
O              & 24i        & 0.57        & 0.87        & 0.15        \\\hline\hline              
\end{tabular}
\end{table}

As we discussed earlier, we start by removing the halide anions from the parental crystal structures. Since Na$_{4}$(AlSiO$_{4}$)$_{3}$Cl and Na$_{4}$(AlSiO$_{4}$)$_{3}$Br share the same framework, this lead to three candidate structures for further consideration. After the structural relaxation, we found that there exist notable imaginary frequencies in the computed phonon spectrum of Na$_{4}$(AlSiO$_{4}$)$_{3}$ and Na$_{4}$(AlGeO$_{4}$)$_{3}$. As shown in the left panel of Fig. ~\ref{Fig_phonon}a for Na$_{4}$(AlSiO$_{4}$)$_{3}$, the most negative phonon frequency is located at the $\Gamma$ point. Therefore, We derived a new structure by displacing atoms along the eigenvector of the soft mode (see the middle panel of Fig.~\ref{Fig_phonon}a) with the maximum magnitude of 0.2 $\textrm{\AA}$ on the Na atoms. Then, the new structure was relaxed again. We repeated this process until the absolute minimum of the potential energy surface was reached and no imaginary frequency phonon occurred. The similar method has been used by the authors in a previous study \cite{kang_prb2014}. After these treatments, we obtained new structures with the energy drops of 71 and 4 meV per formula unit for Na$_{4}$(AlSiO$_{4}$)$_{3}$ and Na$_{4}$(AlGeO$_{4}$)$_{3}$, respectively. On the other hand, Na$_{4}$(GaSiO$_{4}$)$_{3}$ retains the same space group symmetry as found in the original sodalite form.

Due to the soft modes, the original Na (8e) and O (24i) sites in space group $P\bar{4}3n$ were split into the 4e and 12j sites in space group $P23$. Therefore, the final stable structure for these two electrides has a subgroup symmetry $P$23 (195).  Table \ref{tab_xtal} summarizes the crystallographic information for all three structures. Compared to Table \ref{tab_pro}, one can clearly found that the cell parameters nearly remains the same regardless of either the symmetry change or atomic removal.

\subsection{Electronic Properties}

\begin{figure*}[htbp]
\centering
\includegraphics[width=0.98 \textwidth]{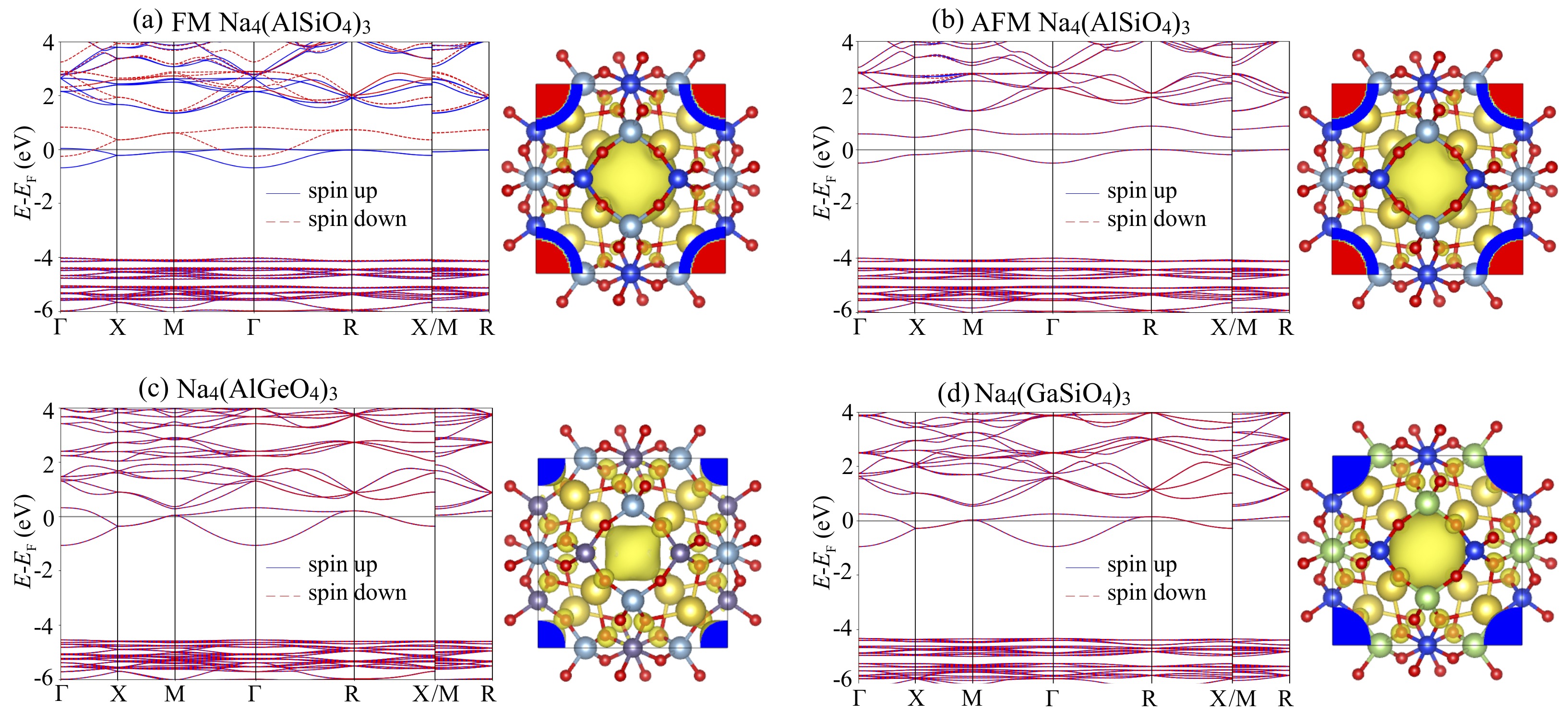}
\caption{\label{Fig_band} The electronic band structures and isosurfaces of partial charge density in the range of -2 $<$ $E$ - $E_{\mathrm{F}}$ $<$ 0 eV of (a) ferromagnetic Na$_{4}$(AlSiO$_{4}$)$_{3}$, (b) antiferromagnetic Na$_{4}$(AlSiO$_{4}$)$_{3}$, (c) Na$_{4}$(AlGeO$_{4}$)$_{3}$, and (d) Na$_{4}$(GaSiO$_{4}$)$_{3}$.
In each band structure, blue solid lines represent spin up bands and red dashed lines represent spin down bands.}
\end{figure*}

Next, we investigated the electronic properties of the resulting three crystalline electrides.
In particular, the Cl removed electride Na$_{4}$(AlSiO$_{4}$)$_{3}$ corresponds to the previously investigated black sodalite, in which each periodic $\beta$-cage contains one Na$_{4}$$^{3+}$ cluster and its color is black due to a metallic band structure \cite{srdanov_jpc1992,smeulders_zeolites1987,nick_jcp1996}. 
As shown in Fig.~\ref{Fig_band}, the three structures manifest electride bands inside the gap of the parent sodalite.
The partial charge density obtained in the energy range $-2 < E-E_{\mathrm{F}} <0 $ is centered inside the cage after the removal of halide anions, indicating that the excess electrons are localized at the crystal cages and forming the electride bands.
There are sizable contributions of oxygen orbitals in the charge density, giving rise to relatively dispersed electride bands in comparison to the valence bands of the parent compounds.
Therefore, the dispersion of the electride bands can be attributed to the hybridization between localized excess electrons in Na$_{4}$ tetrahedron and oxygen orbitals.

\begin{figure*}[htbp]
\centering
\includegraphics[width=0.98 \textwidth]{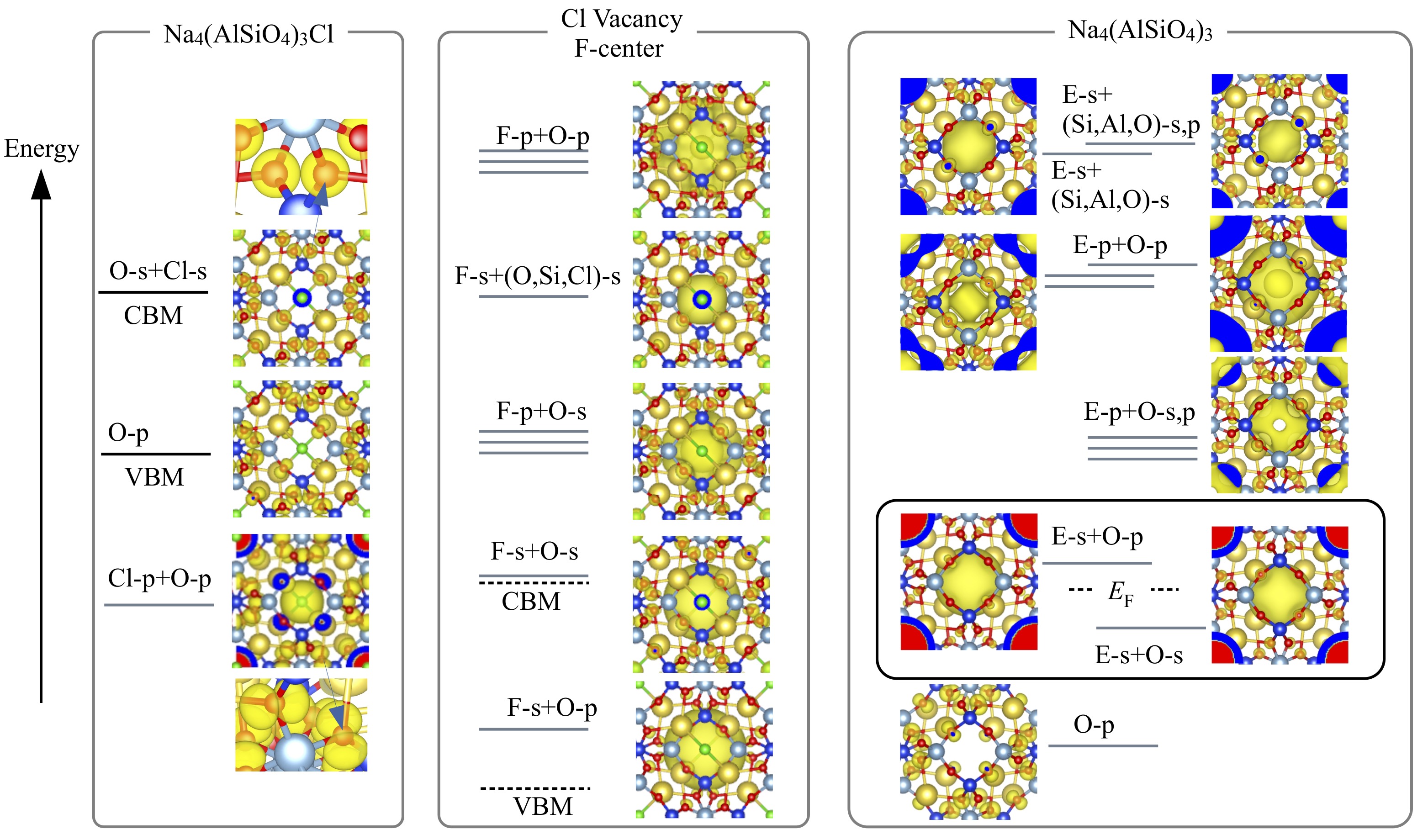}
\caption{\label{Fig_wave} The computed wavefunctions for pristine Na$_{4}$(AlSiO$_{4}$)$_{3}$Cl, F-center of Cl vacancy in the supercell of Na$_{4}$(AlSiO$_{4}$)$_{3}$Cl, and electride states in Na$_{4}$(AlSiO$_{4}$)$_{3}$.
F and E denote the F-center and electride state, respectively.
In the left panel, the magnified views of O-$p$ and O-$s$ wavefunctions are displayed to aid the visual analysis.
Three gray lines denote states that are triply degenerate.
In the right panel, wavefunctions which consist of electride bands are marked by black boxes.}
\end{figure*}

%Our analyses agree with that at the highest loading of Na defects in sodalite cage, the color centers are arranged in a body-centered cubic lattice, allowing the electrons associated with the centers to form bands. This may explain the black color observed at high concentration.\cite{nick_jcp1996}

%Our analyses agree with Prolonged exposure of colorless dry sodalite to alkali vapor causes the material to gradually turn blue, dark blue, and finally black. The blue color observed at low sodium uptake appears because the absorbed sodium atoms are spontaneously ionized. The electron produced by ionization is shared by the four sodium ions present in the sodalite cage three initially there and the fourth originating from the absorbed atom!. The color center created in this way is represented by the formula ~Na 1 ! 4 eF 32 . Here, e stands for the electron and F 32 for the negatively charged frame surrounding a zeolite cage. At the highest loading, when each cage contains an absorbed alkali atom, the color centers are arranged in a body-centered cubic lattice, allowing the electrons associated with the centers to form bands. This may explain the black color observed at high concentration.\cite{nick_jcp1996}

Recently, Stoliaroff et al. have suggested that high concentrations of chlorine vacancies can be created by F-center point defects using DFT modeling \cite{adreien_jpcc2021}. 
It was found that ground and excited states of the singly occupied molecular orbital and the triply degenerate lowest unoccupied molecular orbital for the F-center \cite{pauline_jpcc2020}.
It was observed that the F-center is an electron trapped at an halide anion vacancy in alkali halides \cite{holton_prb1962}.
The trapped electron is in a set of quantized states similar to a hydrogen atom: the ground state is the atomic 1$s$ state, and the first excited state is composed of the 2$s$ and 2$p$ states \cite{takeshi_rpp2011}. 
However, it is shown that hybridization between metal and electride orbitals leads to large splitting between electride associated states \cite{lauren_jacs2022}.
Electride associated states are split into two when hybridization between metal and electride orbitals occurs.
Otto et al. proposed that the particle-in-a-box state is better described as a particle in a box hybridized with framework oxygen atoms for the black sodalite Na$_{4}$(AlSiO$_{4}$)$_{3}$ \cite{ott_prb1998}.
In this model, the particle is confined to a box defined by the oxygen atoms, and the electron density is spread more uniformly throughout the structure.
These results suggest that the Cl removed electride can be described by periodic F-center and it may keep dual nature of localized and itinerant electrons.
However, all aforementioned studies on the black sodalite were based on the dynamically unstable $P\bar{4}3n$ Na$_{4}$(AlSiO$_{4}$)$_{3}$.
Thus, we need to investigate the co-operative localization and hybridization in the ground state $P$23 Na$_{4}$(AlSiO$_{4}$)$_{3}$ electride. 

Fig.~\ref{Fig_wave} displays the computed wavefunctions for pristine $P\bar{4}3n$ Na$_{4}$(AlSiO$_{4}$)$_{3}$Cl, F-center of Cl vacancy in the supercell structure, and crystalline $P$23 Na$_{4}$(AlSiO$_{4}$)$_{3}$. 
In the pristine structure (see the left panel of Fig. ~\ref{Fig_wave}), it shows that the wavefunction of VBM is comprised of O-$p$ while the CBM is comprised of [O-$s$ + Cl-$s$]. 
The valence band right below the VBM is comprised of [Cl-$p$ + O-$p$], in which the wavefunction manifests a spherical shape inside of Na$_{4}$ tetrahedron at the vicinity of Cl. 
This is a typical for an ionic bonding in compound like NaCl.

As shown in the middle panel of Fig. ~\ref{Fig_wave}, the introduction of Cl vacancy gives rise to the formation of a defect F-center. 
Like the atomic orbital, the F center exhibits a set of quantized states similar to a hydrogen atom. The first F-center associated state appears between the VBM and CBM. 
In comparison to the spherical [Cl-$p$ + O-$p$] state in the pristine structure, this state is mainly comprised of localized F-$s$ state in the Na$_{4}$ tetrahedron and a few O-$p$ states centered at the O atoms.
Around the CBM, we observed the hybridization of [F-$s$ + O-$p$] as the first unoccupied state. 
At the second lowest conduction bands, we also found triply degenerate hybridization of F-$p$ and O-$s$, where a spherical wavefunction inside Na$_{4}$ tetrahedron contains an empty region at the center. Similar to F-$s$, the linear combination of F-$p$ also has a spherical shape, but it has an inner hollow due to the requirement of orthogonality. Furthermore, the higher energy conduction bands consist of [F-$s$ + (O, Si, Cl)-$s$] and another triply degenerate [F-$p$ + O-$p$].

Finally, the crystalline electride states (as shown in the right panel of Fig.~\ref{Fig_wave}) have qualitatively the same properties with F-center. Their electride states exhibit distinct local $s$- or $p$-like features similar to F-$s$ and F-$p$. These E-$s$ and E-$p$ states also have notable hybridization with the surrounding atomic orbitals. Due to the hybridization, the band width becomes larger, suggesting a notable excess electron hopping through $\beta$ frames in the crystalline electride \cite{ott_prb1998}.
%Our results are consistent with the previous analysis \cite{ott_prb1998}.

\subsection{Magnetism of Na$_{4}$(AlSiO$_{4}$)$_{3}$}

\begin{figure*}[ht]
\centering
\includegraphics[width=0.99 \textwidth]{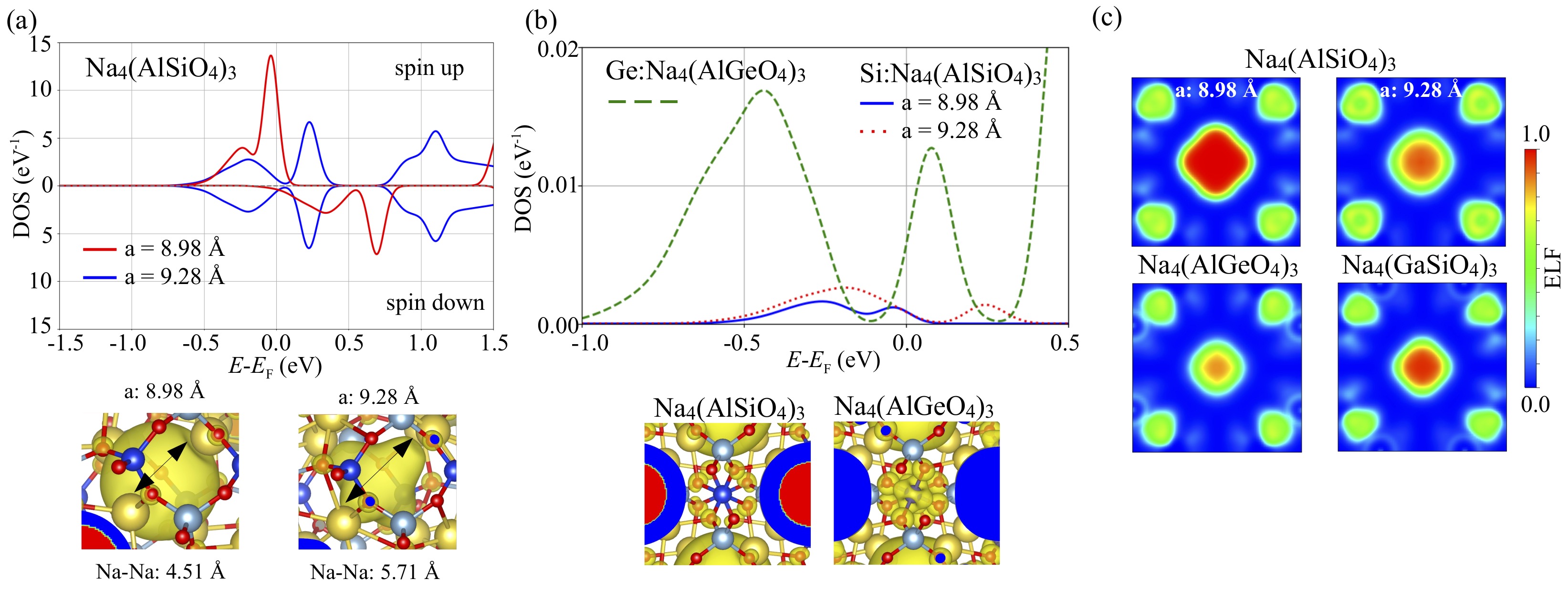}
\caption{\label{Fig_dos} The calculated electronic structures of electrides.
(a) The total density of states and wavefunctions of fully occupied band at $\Gamma$ point for Na$_4$(AlSiO$_4$)$_3$ with two different lattice parameters.
(b) Atomic (Ge,Si) projected density of states of Na$_{4}$(AlGeO$_{4}$)$_{3}$, Na$_{4}$(AlSiO$_{4}$)$_{3}$ (a=8.98 $\textrm{\AA}$), and Na$_{4}$(AlSiO$_{4}$)$_{3}$ (a=9.28 $\textrm{\AA}$).
The spin up density of states are illustrated.
The bottom panel shows isosurfaces of partial charge density in the range of -2 $<$ $E$ - $E_{\mathrm{F}}$ $<$ 0 eV for Na$_{4}$(AlSiO$_{4}$)$_{3}$ (a=8.98 $\textrm{\AA}$) and Na$_{4}$(AlGeO$_{4}$)$_{3}$.
(c) The electron localization functional maps at (100) plane of Na$_{4}$(AlSiO$_{4}$)$_{3}$ (a=8.98 $\textrm{\AA}$), Na$_{4}$(AlSiO$_{4}$)$_{3}$ (a=9.28 $\textrm{\AA}$), Na$_{4}$(AlGeO$_{4}$)$_{3}$, and Na$_{4}$(GaSiO$_{4}$)$_{3}$.}
\end{figure*}

Of the three proposed electrides, Na$_4$(AlSiO$_4$)$_3$ manifests ferromagnetic (FM) electronic structure. 
Blake et al. first showed that the antiferromagnetic (AFM) transition is associated with the incompletely doped regions of the sample \cite{nick_jcp1996}. But no such transition was observed in the fully doped sample.
However, a later experiment by Heinmaa and Lippmass reported the AFM phase with a Neel temperature of $T_{\mathrm{N}}$ = 54 K for heavily sodium-doped sodalite Na$_{4}$(AlSiO$_{4}$)$_{3}$ via nuclear magnetic resonance spectroscopy study \cite{heinmaa_cpl2000}.
This experiment was also supported by following theoretical calculations. Sankey et al. reported the magnetic ground state of Na$_{4}$(AlSiO$_{4}$)$_{3}$ (i.e. fully doped case) by conducting local-spin-density approximation (LSDA) simulations \cite{ott_prb1998}. They showed a lower energy of AFM ordering was favored over the FM ordering.
Our calculations also reveal that AFM configuration has a lower energy of 33 meV per formula unit than FM configuration. 
As shown in Fig. \ref{Fig_band}b, the AFM spin ordering opens a gap, which is an indirect gap of 0.45 eV from R to X point, which is consistent with the previous study on the $P\bar{4}3n$ (Na, K)$_4$(AlSiO$_4$)$_3$ \cite{madsen2001electronic}.

In this work, we also investigated the origin of magnetization within FM configuration to compare existing ferromagnetic electrides Y$_{2}$C \cite{Y2C-PRB-2015} and Ca$_{5}$Ga$_{2}$N$_{4}$ \cite{wang2018ternary}. 
The calculated total energy of spin-polarized Na$_{4}$(AlSiO$_{4}$)$_{3}$ is lower by 42 meV per formula unit than the non-spin-polarized system. The total magnetization moment is 0.97 $\mu_{\textrm{B}}$ per formula unit. These values are between $\sim$ 15 meV and 0.38 $\mu_{\textrm{B}}$ for a weak itinerant ferromagnetic electride Y$_{2}$C \cite{Y2C-PRB-2015}, and 65 meV and 1.25 $\mu_{\textrm{B}}$ for Ca$_{5}$Ga$_{2}$N$_{4}$ \cite{wang2018ternary}. 

%We demonstrated that this magnetism originates from a highly localized excess electron state surrounded by electron-positive alkaline cations.
Similar to the previous study \cite{wang2018ternary}, we found that the strong magnetization of Na$_{4}$(AlSiO$_{4}$)$_{3}$ is due to excess electron localization inside the Na$_{4}$ tetrahedron.
As shown in Fig. \ref{Fig_dos}a, increasing lattice parameter (length of Na$_{4}$ tetrahedron edges) from 8.98 $\textrm{\AA}$ (4.51 $\textrm{\AA}$) to 9.28 $\textrm{\AA}$ (5.71 $\textrm{\AA}$) results in less excess electron confinement in the larger Na$_{4}$ tetrahedron making the magnetic moments disappear. 
The electron localization can be determined by the electron localization function (ELF), which is a measure of the likelihood of finding an electron in the neighborhood space of a reference electron located at a given point \cite{ELF}. The ELF value is a dimensionless index. ELF = 1 corresponding to perfect localization and ELF = $\frac{1}{2}$ corresponding to the electron gas.
As shown in Fig. \ref{Fig_dos}c, the maximum ELF value at the cage center for Na$_{4}$(AlSiO$_{4}$)$_{3}$ is 0.99 when a=8.98 $\textrm{\AA}$, which is higher than 0.94 when a=9.28 $\textrm{\AA}$. For the reference, the corresponding maximum ELF values are 0.85 for Na$_{4}$(AlGeO$_{4}$)$_{3}$, and 0.96 for Na$_{4}$(GaSiO$_{4}$)$_{3}$. The trend of ELF values is also consistent with the results on the band width.
The calculated electride spin up band widths are 0.72 eV for Na$_{4}$(AlSiO$_{4}$)$_{3}$ with a=8.98 $\textrm{\AA}$), 1.19 eV for Na$_{4}$(AlSiO$_{4}$)$_{3}$ with a=9.28 $\textrm{\AA}$, 1.39 eV for Na$_{4}$(AlGeO$_{4}$)$_{3}$, and 1.20 eV for Na$_{4}$(GaSiO$_{4}$)$_{3}$.
Clearly, these localizations are strongly associate with the electride band widths.

For itinerant ferromagnets, magnetism can be understood by the Stoner criterion of $D(E_{\mathrm{F}})I > 1$, where $D(E_{\mathrm{F}})$ is density of state at the Fermi level and $I$ is the exchange parameter. As shown in Fig. \ref{Fig_dos}a, the sharp $D(E_{\mathrm{F}})$ of Na$_{4}$(AlSiO$_{4}$)$_{3}$ (a=8.98 $\textrm{\AA}$) suggests the magnetism is raised from the Stoner-type instability. 
We showed the sharp $D(E_{\mathrm{F}})$, which associates with the small electride band width, is due to localization of excess electron.  
Then, a question is raised for the non-magnetism of Na$_{4}$(AlSiO$_{4}$)$_{3}$ and Na$_{4}$(GaSiO$_{4}$)$_{3}$. 
We checked the possible magnetism of Na$_{4}$(AlGeO$_{4}$)$_{3}$ with smaller lattice parameters to try to localize the excess electron and increase $D(E_{\mathrm{F}})$. All attempts resulted in non-spin-polarized electronic structures.
This can be attributed to the higher electronegativity of Ge than Si.
As shown in Fig. \ref{Fig_dos}b, owing to the electronegativity, the Ge's projected DOS (PDOS) in Na$_{4}$(AlGeO$_{4}$)$_{3}$ is greater than the Si's PDOS in Na$_{4}$(GaSiO$_{4}$)$_{3}$ below the Fermi level, while there is no notable difference in the Si's PDOS in the two different lattices.
The large metal cation's electronegativity leads to increasing hybridization \cite{lauren_jacs2022}, which results in less localization of excess electrons and a large band width. This hinders spin polarization of the excess electrons. Therefore, there is no magnetism in either Na$_{4}$(AlSiO$_{4}$)$_{3}$ or Na$_{4}$(GaSiO$_{4}$)$_{3}$. 

%\clearpage
\section{SUMMARY AND CONCLUSIONS}
In sum, we present an approach to realize the electride material via the crystallographic symmetry analysis. Inspired by the synthesis of C12A7:2e$^-$, we proposed to make potentially stable electrides from thermally stable inorganic compounds that possess the high symmetry Wyckoff site being occupied by anions. Using the well-known halide sodalites as the examples, we found that the materials after the removal of anionic halide anions exhibit typical electride behaviors that are characterized by the existence of localized electronic states near the Fermi level. Among them, Na$_4$(AlSiO$_4$)$_3$ manifests antiferromagnetic electronic structure. We demonstrate that this magnetism originate from a highly localized excess electron state surrounded by electronpositive alkaline cations. While some compounds (e.g.,  Na$_4$(AlSiO$_4$)$_3$) have been investigated previously due to interest in F-center research \cite{nick_jcp1996, Srdanov-PRL-1998, ott_prb1998, heinmaa_cpl2000, madsen2001electronic}, our work provided more rigorous phonon analysis to derive the truly stable ground state structures of crystalline electrides. Compared to most previously studied electrides, these materials are derived from the sodalite materials with high melting points. Therefore, they are expected to be more thermally stable due to the complex structural framework. Thanks to the improved stability, these predicted materials are more amenable for fabrication and practical applications. In the present work, we only investigated several halide sodalite to demonstrate the proof of concept. The extension to other sodalite type materials with different anions is the subject of future work. In addition, our work suggests the connection between porous zeolite and electride since the existence of crystalline zeolite can naturally provide the crystal cavity to accommodate the excess electrons. While efforts on making zeolite electrides have been attempted in the past \cite{Petkov-PRL-2002, wernette2003inorganic}, the extraordinary diversity in zeolite structures \cite{zeo} warrants many opportunities for future exploration. 

\section*{Acknowledgments}
This research was sponsored by the U.S. Department of Energy, Office of Science, Office of Basic Energy Sciences, Theoretical Condensed Matter Physics program and the DOE Established Program to Stimulate Competitive Research under Award Number DE-SC0021970. The computing resources are provided by ACCESS (TG-DMR180040) and National Energy Research Scientific Computing Center (NERSC).

\bibliography{ref.bib}

\begin{thebibliography}{72}%
\makeatletter
\providecommand \@ifxundefined [1]{%
 \@ifx{#1\undefined}
}%
\providecommand \@ifnum [1]{%
 \ifnum #1\expandafter \@firstoftwo
 \else \expandafter \@secondoftwo
 \fi
}%
\providecommand \@ifx [1]{%
 \ifx #1\expandafter \@firstoftwo
 \else \expandafter \@secondoftwo
 \fi
}%
\providecommand \natexlab [1]{#1}%
\providecommand \enquote  [1]{``#1''}%
\providecommand \bibnamefont  [1]{#1}%
\providecommand \bibfnamefont [1]{#1}%
\providecommand \citenamefont [1]{#1}%
\providecommand \href@noop [0]{\@secondoftwo}%
\providecommand \href [0]{\begingroup \@sanitize@url \@href}%
\providecommand \@href[1]{\@@startlink{#1}\@@href}%
\providecommand \@@href[1]{\endgroup#1\@@endlink}%
\providecommand \@sanitize@url [0]{\catcode `\\12\catcode `\$12\catcode
  `\&12\catcode `\#12\catcode `\^12\catcode `\_12\catcode `\%12\relax}%
\providecommand \@@startlink[1]{}%
\providecommand \@@endlink[0]{}%
\providecommand \url  [0]{\begingroup\@sanitize@url \@url }%
\providecommand \@url [1]{\endgroup\@href {#1}{\urlprefix }}%
\providecommand \urlprefix  [0]{URL }%
\providecommand \Eprint [0]{\href }%
\providecommand \doibase [0]{http://dx.doi.org/}%
\providecommand \selectlanguage [0]{\@gobble}%
\providecommand \bibinfo  [0]{\@secondoftwo}%
\providecommand \bibfield  [0]{\@secondoftwo}%
\providecommand \translation [1]{[#1]}%
\providecommand \BibitemOpen [0]{}%
\providecommand \bibitemStop [0]{}%
\providecommand \bibitemNoStop [0]{.\EOS\space}%
\providecommand \EOS [0]{\spacefactor3000\relax}%
\providecommand \BibitemShut  [1]{\csname bibitem#1\endcsname}%
\let\auto@bib@innerbib\@empty
%</preamble>
\bibitem [{\citenamefont {Dye}(1990)}]{Dye-Science-1990}%
  \BibitemOpen
  \bibfield  {author} {\bibinfo {author} {\bibfnamefont {J.~L.}\ \bibnamefont
  {Dye}},\ }\href {\doibase 10.1126/science.247.4943.663} {\bibfield  {journal}
  {\bibinfo  {journal} {Science}\ }\textbf {\bibinfo {volume} {247}},\ \bibinfo
  {pages} {663} (\bibinfo {year} {1990})}\BibitemShut {NoStop}%
\bibitem [{\citenamefont {Li}\ and\ \citenamefont
  {Mahanti}(2004)}]{li2004theoretical}%
  \BibitemOpen
  \bibfield  {author} {\bibinfo {author} {\bibfnamefont {H.}~\bibnamefont
  {Li}}\ and\ \bibinfo {author} {\bibfnamefont {S.}~\bibnamefont {Mahanti}},\
  }\href {\doibase 10.1103/PhysRevLett.93.216406} {\bibfield  {journal}
  {\bibinfo  {journal} {Phys. Rev. Lett.}\ }\textbf {\bibinfo {volume} {93}},\
  \bibinfo {pages} {216406} (\bibinfo {year} {2004})}\BibitemShut {NoStop}%
\bibitem [{\citenamefont {Dye}(2009)}]{dye2009electrides}%
  \BibitemOpen
  \bibfield  {author} {\bibinfo {author} {\bibfnamefont {J.~L.}\ \bibnamefont
  {Dye}},\ }\href {\doibase 10.1021/ar9000857} {\bibfield  {journal} {\bibinfo
  {journal} {Acc. Chem. Res.}\ }\textbf {\bibinfo {volume} {42}},\ \bibinfo
  {pages} {1564} (\bibinfo {year} {2009})}\BibitemShut {NoStop}%
\bibitem [{\citenamefont {Hosono}\ and\ \citenamefont
  {Kitano}(2021)}]{hosono2021advances}%
  \BibitemOpen
  \bibfield  {author} {\bibinfo {author} {\bibfnamefont {H.}~\bibnamefont
  {Hosono}}\ and\ \bibinfo {author} {\bibfnamefont {M.}~\bibnamefont
  {Kitano}},\ }\href {\doibase 10.1021/acs.chemrev.0c01071} {\bibfield
  {journal} {\bibinfo  {journal} {Chem. Rev.}\ }\textbf {\bibinfo {volume}
  {121}},\ \bibinfo {pages} {3121} (\bibinfo {year} {2021})}\BibitemShut
  {NoStop}%
\bibitem [{\citenamefont {Liu}\ \emph {et~al.}(2020)\citenamefont {Liu},
  \citenamefont {Nikolaev}, \citenamefont {Ren},\ and\ \citenamefont
  {Burton}}]{liu2020electrides}%
  \BibitemOpen
  \bibfield  {author} {\bibinfo {author} {\bibfnamefont {C.}~\bibnamefont
  {Liu}}, \bibinfo {author} {\bibfnamefont {S.~A.}\ \bibnamefont {Nikolaev}},
  \bibinfo {author} {\bibfnamefont {W.}~\bibnamefont {Ren}}, \ and\ \bibinfo
  {author} {\bibfnamefont {L.~A.}\ \bibnamefont {Burton}},\ }\href {\doibase
  10.1039/D0TC01165G} {\bibfield  {journal} {\bibinfo  {journal} {J. Mater.
  Chem. C}\ }\textbf {\bibinfo {volume} {8}},\ \bibinfo {pages} {10551}
  (\bibinfo {year} {2020})}\BibitemShut {NoStop}%
\bibitem [{\citenamefont {Inoshita}\ \emph {et~al.}(2021)\citenamefont
  {Inoshita}, \citenamefont {Saito},\ and\ \citenamefont
  {Hosono}}]{inoshita2021floating}%
  \BibitemOpen
  \bibfield  {author} {\bibinfo {author} {\bibfnamefont {T.}~\bibnamefont
  {Inoshita}}, \bibinfo {author} {\bibfnamefont {S.}~\bibnamefont {Saito}}, \
  and\ \bibinfo {author} {\bibfnamefont {H.}~\bibnamefont {Hosono}},\ }\href
  {\doibase 10.1002/smsc.202100020} {\bibfield  {journal} {\bibinfo  {journal}
  {Small Science}\ }\textbf {\bibinfo {volume} {1}},\ \bibinfo {pages}
  {2100020} (\bibinfo {year} {2021})}\BibitemShut {NoStop}%
\bibitem [{\citenamefont {Ellaboudy}\ \emph {et~al.}(1983)\citenamefont
  {Ellaboudy}, \citenamefont {Dye},\ and\ \citenamefont
  {Smith}}]{Ellaboudy-1983-JACS}%
  \BibitemOpen
  \bibfield  {author} {\bibinfo {author} {\bibfnamefont {A.}~\bibnamefont
  {Ellaboudy}}, \bibinfo {author} {\bibfnamefont {J.~L.}\ \bibnamefont {Dye}},
  \ and\ \bibinfo {author} {\bibfnamefont {P.~B.}\ \bibnamefont {Smith}},\
  }\href {\doibase 10.1021/ja00359a022} {\bibfield  {journal} {\bibinfo
  {journal} {J. Am. Chem. Soc.}\ }\textbf {\bibinfo {volume} {105}},\ \bibinfo
  {pages} {6490} (\bibinfo {year} {1983})}\BibitemShut {NoStop}%
\bibitem [{\citenamefont {Matsuishi}\ \emph {et~al.}(2003)\citenamefont
  {Matsuishi}, \citenamefont {Toda}, \citenamefont {Miyakawa}, \citenamefont
  {Hayashi}, \citenamefont {Kamiya}, \citenamefont {Hirano}, \citenamefont
  {Tanaka},\ and\ \citenamefont {Hosono}}]{Matsuishi-Science-2003}%
  \BibitemOpen
  \bibfield  {author} {\bibinfo {author} {\bibfnamefont {S.}~\bibnamefont
  {Matsuishi}}, \bibinfo {author} {\bibfnamefont {Y.}~\bibnamefont {Toda}},
  \bibinfo {author} {\bibfnamefont {M.}~\bibnamefont {Miyakawa}}, \bibinfo
  {author} {\bibfnamefont {K.}~\bibnamefont {Hayashi}}, \bibinfo {author}
  {\bibfnamefont {T.}~\bibnamefont {Kamiya}}, \bibinfo {author} {\bibfnamefont
  {M.}~\bibnamefont {Hirano}}, \bibinfo {author} {\bibfnamefont
  {I.}~\bibnamefont {Tanaka}}, \ and\ \bibinfo {author} {\bibfnamefont
  {H.}~\bibnamefont {Hosono}},\ }\href {\doibase 10.1126/science.1083842}
  {\bibfield  {journal} {\bibinfo  {journal} {Science}\ }\textbf {\bibinfo
  {volume} {301}},\ \bibinfo {pages} {626} (\bibinfo {year}
  {2003})}\BibitemShut {NoStop}%
\bibitem [{\citenamefont {Lee}\ \emph {et~al.}(2013)\citenamefont {Lee},
  \citenamefont {Kim}, \citenamefont {Toda}, \citenamefont {Matsuishi},\ and\
  \citenamefont {Hosono}}]{Lee-Nature-2013}%
  \BibitemOpen
  \bibfield  {author} {\bibinfo {author} {\bibfnamefont {K.}~\bibnamefont
  {Lee}}, \bibinfo {author} {\bibfnamefont {S.~W.}\ \bibnamefont {Kim}},
  \bibinfo {author} {\bibfnamefont {Y.}~\bibnamefont {Toda}}, \bibinfo {author}
  {\bibfnamefont {S.}~\bibnamefont {Matsuishi}}, \ and\ \bibinfo {author}
  {\bibfnamefont {H.}~\bibnamefont {Hosono}},\ }\href {\doibase
  10.1038/nature11812} {\bibfield  {journal} {\bibinfo  {journal} {Nature}\
  }\textbf {\bibinfo {volume} {494}},\ \bibinfo {pages} {336} (\bibinfo {year}
  {2013})}\BibitemShut {NoStop}%
\bibitem [{\citenamefont {Zhang}\ \emph {et~al.}(2014)\citenamefont {Zhang},
  \citenamefont {Xiao}, \citenamefont {Lei}, \citenamefont {Toda},
  \citenamefont {Matsuishi}, \citenamefont {Kamiya}, \citenamefont {Ueda},\
  and\ \citenamefont {Hosono}}]{Zhang-CM-Y2C-2014}%
  \BibitemOpen
  \bibfield  {author} {\bibinfo {author} {\bibfnamefont {X.}~\bibnamefont
  {Zhang}}, \bibinfo {author} {\bibfnamefont {Z.}~\bibnamefont {Xiao}},
  \bibinfo {author} {\bibfnamefont {H.}~\bibnamefont {Lei}}, \bibinfo {author}
  {\bibfnamefont {Y.}~\bibnamefont {Toda}}, \bibinfo {author} {\bibfnamefont
  {S.}~\bibnamefont {Matsuishi}}, \bibinfo {author} {\bibfnamefont
  {T.}~\bibnamefont {Kamiya}}, \bibinfo {author} {\bibfnamefont
  {S.}~\bibnamefont {Ueda}}, \ and\ \bibinfo {author} {\bibfnamefont
  {H.}~\bibnamefont {Hosono}},\ }\href {\doibase 10.1021/cm503512h} {\bibfield
  {journal} {\bibinfo  {journal} {Chem. Mater.}\ }\textbf {\bibinfo {volume}
  {26}},\ \bibinfo {pages} {6638} (\bibinfo {year} {2014})}\BibitemShut
  {NoStop}%
\bibitem [{\citenamefont {Lu}\ \emph {et~al.}(2016)\citenamefont {Lu},
  \citenamefont {Li}, \citenamefont {Tada}, \citenamefont {Toda}, \citenamefont
  {Ueda}, \citenamefont {Yokoyama}, \citenamefont {Kitano},\ and\ \citenamefont
  {Hosono}}]{Lu-JACS-2016}%
  \BibitemOpen
  \bibfield  {author} {\bibinfo {author} {\bibfnamefont {Y.}~\bibnamefont
  {Lu}}, \bibinfo {author} {\bibfnamefont {J.}~\bibnamefont {Li}}, \bibinfo
  {author} {\bibfnamefont {T.}~\bibnamefont {Tada}}, \bibinfo {author}
  {\bibfnamefont {Y.}~\bibnamefont {Toda}}, \bibinfo {author} {\bibfnamefont
  {S.}~\bibnamefont {Ueda}}, \bibinfo {author} {\bibfnamefont {T.}~\bibnamefont
  {Yokoyama}}, \bibinfo {author} {\bibfnamefont {M.}~\bibnamefont {Kitano}}, \
  and\ \bibinfo {author} {\bibfnamefont {H.}~\bibnamefont {Hosono}},\ }\href
  {\doibase 10.1021/jacs.6b00124} {\bibfield  {journal} {\bibinfo  {journal}
  {J. Am. Chem. Soc.}\ }\textbf {\bibinfo {volume} {138}},\ \bibinfo {pages}
  {3970} (\bibinfo {year} {2016})}\BibitemShut {NoStop}%
\bibitem [{\citenamefont {Wang}\ \emph {et~al.}(2017)\citenamefont {Wang},
  \citenamefont {Hanzawa}, \citenamefont {Hiramatsu}, \citenamefont {Kim},
  \citenamefont {Umezawa}, \citenamefont {Iwanaka}, \citenamefont {Tada},\ and\
  \citenamefont {Hosono}}]{Wang-JACS-2017}%
  \BibitemOpen
  \bibfield  {author} {\bibinfo {author} {\bibfnamefont {J.}~\bibnamefont
  {Wang}}, \bibinfo {author} {\bibfnamefont {K.}~\bibnamefont {Hanzawa}},
  \bibinfo {author} {\bibfnamefont {H.}~\bibnamefont {Hiramatsu}}, \bibinfo
  {author} {\bibfnamefont {J.}~\bibnamefont {Kim}}, \bibinfo {author}
  {\bibfnamefont {N.}~\bibnamefont {Umezawa}}, \bibinfo {author} {\bibfnamefont
  {K.}~\bibnamefont {Iwanaka}}, \bibinfo {author} {\bibfnamefont
  {T.}~\bibnamefont {Tada}}, \ and\ \bibinfo {author} {\bibfnamefont
  {H.}~\bibnamefont {Hosono}},\ }\href {\doibase 10.1021/jacs.7b06279}
  {\bibfield  {journal} {\bibinfo  {journal} {J. Am. Chem. Soc.}\ }\textbf
  {\bibinfo {volume} {139}},\ \bibinfo {pages} {15668} (\bibinfo {year}
  {2017})}\BibitemShut {NoStop}%
\bibitem [{\citenamefont {Zhang}\ \emph
  {et~al.}(2017{\natexlab{a}})\citenamefont {Zhang}, \citenamefont {Wang},
  \citenamefont {Xiao}, \citenamefont {Lu}, \citenamefont {Kamiya},
  \citenamefont {Uwatoko}, \citenamefont {Kageyama},\ and\ \citenamefont
  {Hosono}}]{Zhang-QM-2017}%
  \BibitemOpen
  \bibfield  {author} {\bibinfo {author} {\bibfnamefont {Y.}~\bibnamefont
  {Zhang}}, \bibinfo {author} {\bibfnamefont {B.}~\bibnamefont {Wang}},
  \bibinfo {author} {\bibfnamefont {Z.}~\bibnamefont {Xiao}}, \bibinfo {author}
  {\bibfnamefont {Y.}~\bibnamefont {Lu}}, \bibinfo {author} {\bibfnamefont
  {T.}~\bibnamefont {Kamiya}}, \bibinfo {author} {\bibfnamefont
  {Y.}~\bibnamefont {Uwatoko}}, \bibinfo {author} {\bibfnamefont
  {H.}~\bibnamefont {Kageyama}}, \ and\ \bibinfo {author} {\bibfnamefont
  {H.}~\bibnamefont {Hosono}},\ }\href {\doibase 10.1038/s41535-017-0053-4}
  {\bibfield  {journal} {\bibinfo  {journal} {npj Quantum Mater.}\ }\textbf
  {\bibinfo {volume} {2}},\ \bibinfo {pages} {45} (\bibinfo {year}
  {2017}{\natexlab{a}})}\BibitemShut {NoStop}%
\bibitem [{\citenamefont {Zhang}\ \emph {et~al.}(2015)\citenamefont {Zhang},
  \citenamefont {Xiao}, \citenamefont {Kamiya},\ and\ \citenamefont
  {Hosono}}]{Zhang-JPCL-2015}%
  \BibitemOpen
  \bibfield  {author} {\bibinfo {author} {\bibfnamefont {Y.}~\bibnamefont
  {Zhang}}, \bibinfo {author} {\bibfnamefont {Z.}~\bibnamefont {Xiao}},
  \bibinfo {author} {\bibfnamefont {T.}~\bibnamefont {Kamiya}}, \ and\ \bibinfo
  {author} {\bibfnamefont {H.}~\bibnamefont {Hosono}},\ }\href {\doibase
  10.1021/acs.jpclett.5b02283} {\bibfield  {journal} {\bibinfo  {journal} {J.
  Phys. Chem. Lett.}\ }\textbf {\bibinfo {volume} {6}},\ \bibinfo {pages}
  {4966} (\bibinfo {year} {2015})}\BibitemShut {NoStop}%
\bibitem [{\citenamefont {Mizoguchi}\ \emph {et~al.}(2016)\citenamefont
  {Mizoguchi}, \citenamefont {Okunaka}, \citenamefont {Kitano}, \citenamefont
  {Matsuishi}, \citenamefont {Yokoyama},\ and\ \citenamefont
  {Hosono}}]{mizoguchi2016hydride}%
  \BibitemOpen
  \bibfield  {author} {\bibinfo {author} {\bibfnamefont {H.}~\bibnamefont
  {Mizoguchi}}, \bibinfo {author} {\bibfnamefont {M.}~\bibnamefont {Okunaka}},
  \bibinfo {author} {\bibfnamefont {M.}~\bibnamefont {Kitano}}, \bibinfo
  {author} {\bibfnamefont {S.}~\bibnamefont {Matsuishi}}, \bibinfo {author}
  {\bibfnamefont {T.}~\bibnamefont {Yokoyama}}, \ and\ \bibinfo {author}
  {\bibfnamefont {H.}~\bibnamefont {Hosono}},\ }\href {\doibase
  10.1021/acs.inorgchem.6b01369} {\bibfield  {journal} {\bibinfo  {journal}
  {Inorg. Chem.}\ }\textbf {\bibinfo {volume} {55}},\ \bibinfo {pages} {8833}
  (\bibinfo {year} {2016})}\BibitemShut {NoStop}%
\bibitem [{\citenamefont {Inoshita}\ \emph {et~al.}(2014)\citenamefont
  {Inoshita}, \citenamefont {Jeong}, \citenamefont {Hamada},\ and\
  \citenamefont {Hosono}}]{Inoshita-PRX-2014}%
  \BibitemOpen
  \bibfield  {author} {\bibinfo {author} {\bibfnamefont {T.}~\bibnamefont
  {Inoshita}}, \bibinfo {author} {\bibfnamefont {S.}~\bibnamefont {Jeong}},
  \bibinfo {author} {\bibfnamefont {N.}~\bibnamefont {Hamada}}, \ and\ \bibinfo
  {author} {\bibfnamefont {H.}~\bibnamefont {Hosono}},\ }\href {\doibase
  10.1103/PhysRevX.4.031023} {\bibfield  {journal} {\bibinfo  {journal} {Phys.
  Rev. X}\ }\textbf {\bibinfo {volume} {4}},\ \bibinfo {pages} {031023}
  (\bibinfo {year} {2014})}\BibitemShut {NoStop}%
\bibitem [{\citenamefont {Inoshita}\ \emph {et~al.}(2015)\citenamefont
  {Inoshita}, \citenamefont {Hamada},\ and\ \citenamefont
  {Hosono}}]{Y2C-PRB-2015}%
  \BibitemOpen
  \bibfield  {author} {\bibinfo {author} {\bibfnamefont {T.}~\bibnamefont
  {Inoshita}}, \bibinfo {author} {\bibfnamefont {N.}~\bibnamefont {Hamada}}, \
  and\ \bibinfo {author} {\bibfnamefont {H.}~\bibnamefont {Hosono}},\ }\href
  {\doibase 10.1103/PhysRevB.92.201109} {\bibfield  {journal} {\bibinfo
  {journal} {Phys. Rev. B}\ }\textbf {\bibinfo {volume} {92}},\ \bibinfo
  {pages} {201109} (\bibinfo {year} {2015})}\BibitemShut {NoStop}%
\bibitem [{\citenamefont {Tada}\ \emph {et~al.}(2014)\citenamefont {Tada},
  \citenamefont {Takemoto}, \citenamefont {Matsuishi},\ and\ \citenamefont
  {Hosono}}]{Tada-IC-2014}%
  \BibitemOpen
  \bibfield  {author} {\bibinfo {author} {\bibfnamefont {T.}~\bibnamefont
  {Tada}}, \bibinfo {author} {\bibfnamefont {S.}~\bibnamefont {Takemoto}},
  \bibinfo {author} {\bibfnamefont {S.}~\bibnamefont {Matsuishi}}, \ and\
  \bibinfo {author} {\bibfnamefont {H.}~\bibnamefont {Hosono}},\ }\href
  {\doibase 10.1021/ic501362b} {\bibfield  {journal} {\bibinfo  {journal}
  {Inorg. Chem.}\ }\textbf {\bibinfo {volume} {53}},\ \bibinfo {pages} {10347}
  (\bibinfo {year} {2014})}\BibitemShut {NoStop}%
\bibitem [{\citenamefont {Ming}\ \emph {et~al.}(2016)\citenamefont {Ming},
  \citenamefont {Yoon}, \citenamefont {Du}, \citenamefont {Lee},\ and\
  \citenamefont {Kim}}]{Ming-JACS-2016}%
  \BibitemOpen
  \bibfield  {author} {\bibinfo {author} {\bibfnamefont {W.}~\bibnamefont
  {Ming}}, \bibinfo {author} {\bibfnamefont {M.}~\bibnamefont {Yoon}}, \bibinfo
  {author} {\bibfnamefont {M.-H.}\ \bibnamefont {Du}}, \bibinfo {author}
  {\bibfnamefont {K.}~\bibnamefont {Lee}}, \ and\ \bibinfo {author}
  {\bibfnamefont {S.~W.}\ \bibnamefont {Kim}},\ }\href {\doibase
  10.1021/jacs.6b05586} {\bibfield  {journal} {\bibinfo  {journal} {J. Am.
  Chem. Soc.}\ }\textbf {\bibinfo {volume} {138}},\ \bibinfo {pages} {15336}
  (\bibinfo {year} {2016})}\BibitemShut {NoStop}%
\bibitem [{\citenamefont {Zhang}\ \emph
  {et~al.}(2017{\natexlab{b}})\citenamefont {Zhang}, \citenamefont {Wang},
  \citenamefont {Wang}, \citenamefont {Zhang},\ and\ \citenamefont
  {Ma}}]{Zhang-PRX-2017}%
  \BibitemOpen
  \bibfield  {author} {\bibinfo {author} {\bibfnamefont {Y.}~\bibnamefont
  {Zhang}}, \bibinfo {author} {\bibfnamefont {H.}~\bibnamefont {Wang}},
  \bibinfo {author} {\bibfnamefont {Y.}~\bibnamefont {Wang}}, \bibinfo {author}
  {\bibfnamefont {L.}~\bibnamefont {Zhang}}, \ and\ \bibinfo {author}
  {\bibfnamefont {Y.}~\bibnamefont {Ma}},\ }\href {\doibase
  10.1103/PhysRevX.7.011017} {\bibfield  {journal} {\bibinfo  {journal} {Phys.
  Rev. X}\ }\textbf {\bibinfo {volume} {7}},\ \bibinfo {pages} {011017}
  (\bibinfo {year} {2017}{\natexlab{b}})}\BibitemShut {NoStop}%
\bibitem [{\citenamefont {Burton}\ \emph {et~al.}(2018)\citenamefont {Burton},
  \citenamefont {Ricci}, \citenamefont {Chen}, \citenamefont {Rignanese},\ and\
  \citenamefont {Hautier}}]{burton2018high}%
  \BibitemOpen
  \bibfield  {author} {\bibinfo {author} {\bibfnamefont {L.~A.}\ \bibnamefont
  {Burton}}, \bibinfo {author} {\bibfnamefont {F.}~\bibnamefont {Ricci}},
  \bibinfo {author} {\bibfnamefont {W.}~\bibnamefont {Chen}}, \bibinfo {author}
  {\bibfnamefont {G.-M.}\ \bibnamefont {Rignanese}}, \ and\ \bibinfo {author}
  {\bibfnamefont {G.}~\bibnamefont {Hautier}},\ }\href {\doibase
  10.1021/acs.chemmater.8b02526} {\bibfield  {journal} {\bibinfo  {journal}
  {Chem. Mater.}\ }\textbf {\bibinfo {volume} {30}},\ \bibinfo {pages} {7521}
  (\bibinfo {year} {2018})}\BibitemShut {NoStop}%
\bibitem [{\citenamefont {Wang}\ \emph {et~al.}(2019)\citenamefont {Wang},
  \citenamefont {Zhu}, \citenamefont {Wang},\ and\ \citenamefont
  {Hosono}}]{wang2018ternary}%
  \BibitemOpen
  \bibfield  {author} {\bibinfo {author} {\bibfnamefont {J.}~\bibnamefont
  {Wang}}, \bibinfo {author} {\bibfnamefont {Q.}~\bibnamefont {Zhu}}, \bibinfo
  {author} {\bibfnamefont {Z.}~\bibnamefont {Wang}}, \ and\ \bibinfo {author}
  {\bibfnamefont {H.}~\bibnamefont {Hosono}},\ }\href {\doibase
  10.1103/PhysRevB.99.064104} {\bibfield  {journal} {\bibinfo  {journal} {Phys.
  Rev. B}\ }\textbf {\bibinfo {volume} {99}},\ \bibinfo {pages} {064104}
  (\bibinfo {year} {2019})}\BibitemShut {NoStop}%
\bibitem [{\citenamefont {Zhu}\ \emph {et~al.}(2019{\natexlab{a}})\citenamefont
  {Zhu}, \citenamefont {Wang}, \citenamefont {Qu}, \citenamefont {Wang},
  \citenamefont {Frolov}, \citenamefont {Chen},\ and\ \citenamefont
  {Zhu}}]{Zhu-PRM-2019}%
  \BibitemOpen
  \bibfield  {author} {\bibinfo {author} {\bibfnamefont {S.-C.}\ \bibnamefont
  {Zhu}}, \bibinfo {author} {\bibfnamefont {L.}~\bibnamefont {Wang}}, \bibinfo
  {author} {\bibfnamefont {J.-Y.}\ \bibnamefont {Qu}}, \bibinfo {author}
  {\bibfnamefont {J.-J.}\ \bibnamefont {Wang}}, \bibinfo {author}
  {\bibfnamefont {T.}~\bibnamefont {Frolov}}, \bibinfo {author} {\bibfnamefont
  {X.-Q.}\ \bibnamefont {Chen}}, \ and\ \bibinfo {author} {\bibfnamefont
  {Q.}~\bibnamefont {Zhu}},\ }\href {\doibase
  10.1103/PhysRevMaterials.3.024205} {\bibfield  {journal} {\bibinfo  {journal}
  {Phys. Rev. Materials}\ }\textbf {\bibinfo {volume} {3}},\ \bibinfo {pages}
  {024205} (\bibinfo {year} {2019}{\natexlab{a}})}\BibitemShut {NoStop}%
\bibitem [{\citenamefont {Qu}\ \emph {et~al.}(2019)\citenamefont {Qu},
  \citenamefont {Zhu}, \citenamefont {Zhang},\ and\ \citenamefont
  {Zhu}}]{Qu-ACSAMI-2019}%
  \BibitemOpen
  \bibfield  {author} {\bibinfo {author} {\bibfnamefont {J.}~\bibnamefont
  {Qu}}, \bibinfo {author} {\bibfnamefont {S.}~\bibnamefont {Zhu}}, \bibinfo
  {author} {\bibfnamefont {W.}~\bibnamefont {Zhang}}, \ and\ \bibinfo {author}
  {\bibfnamefont {Q.}~\bibnamefont {Zhu}},\ }\href {\doibase
  10.1021/acsami.8b18676} {\bibfield  {journal} {\bibinfo  {journal} {ACS Appl.
  Mater. Interfaces}\ }\textbf {\bibinfo {volume} {11}},\ \bibinfo {pages}
  {5256} (\bibinfo {year} {2019})}\BibitemShut {NoStop}%
\bibitem [{\citenamefont {Zhu}\ \emph {et~al.}(2019{\natexlab{b}})\citenamefont
  {Zhu}, \citenamefont {Frolov},\ and\ \citenamefont
  {Choudhary}}]{ZHU20191293}%
  \BibitemOpen
  \bibfield  {author} {\bibinfo {author} {\bibfnamefont {Q.}~\bibnamefont
  {Zhu}}, \bibinfo {author} {\bibfnamefont {T.}~\bibnamefont {Frolov}}, \ and\
  \bibinfo {author} {\bibfnamefont {K.}~\bibnamefont {Choudhary}},\ }\href
  {\doibase https://doi.org/10.1016/j.matt.2019.06.017} {\bibfield  {journal}
  {\bibinfo  {journal} {Matter}\ }\textbf {\bibinfo {volume} {1}},\ \bibinfo
  {pages} {1293 } (\bibinfo {year} {2019}{\natexlab{b}})}\BibitemShut {NoStop}%
\bibitem [{\citenamefont {Li}\ \emph {et~al.}(2021)\citenamefont {Li},
  \citenamefont {Gong}, \citenamefont {Wang},\ and\ \citenamefont
  {Hosono}}]{li2021electron}%
  \BibitemOpen
  \bibfield  {author} {\bibinfo {author} {\bibfnamefont {K.}~\bibnamefont
  {Li}}, \bibinfo {author} {\bibfnamefont {Y.}~\bibnamefont {Gong}}, \bibinfo
  {author} {\bibfnamefont {J.}~\bibnamefont {Wang}}, \ and\ \bibinfo {author}
  {\bibfnamefont {H.}~\bibnamefont {Hosono}},\ }\href {\doibase
  10.1021/jacs.1c03278} {\bibfield  {journal} {\bibinfo  {journal} {J. Am.
  Chem. Soc.}\ }\textbf {\bibinfo {volume} {143}},\ \bibinfo {pages} {8821}
  (\bibinfo {year} {2021})}\BibitemShut {NoStop}%
\bibitem [{\citenamefont {Liu}\ \emph {et~al.}(2022)\citenamefont {Liu},
  \citenamefont {Wei}, \citenamefont {Wang}, \citenamefont {Jiao},\ and\
  \citenamefont {Jing}}]{liu2022preparation}%
  \BibitemOpen
  \bibfield  {author} {\bibinfo {author} {\bibfnamefont {Y.}~\bibnamefont
  {Liu}}, \bibinfo {author} {\bibfnamefont {H.}~\bibnamefont {Wei}}, \bibinfo
  {author} {\bibfnamefont {X.}~\bibnamefont {Wang}}, \bibinfo {author}
  {\bibfnamefont {H.}~\bibnamefont {Jiao}}, \ and\ \bibinfo {author}
  {\bibfnamefont {X.}~\bibnamefont {Jing}},\ }\href {\doibase
  10.1039/D2RA05847B} {\bibfield  {journal} {\bibinfo  {journal} {RSC Adv.}\
  }\textbf {\bibinfo {volume} {12}},\ \bibinfo {pages} {28414} (\bibinfo {year}
  {2022})}\BibitemShut {NoStop}%
\bibitem [{\citenamefont {Yang}\ \emph {et~al.}(2021)\citenamefont {Yang},
  \citenamefont {Parrish}, \citenamefont {Li}, \citenamefont {Sa},
  \citenamefont {Zhan},\ and\ \citenamefont {Zhu}}]{yang-2021-prb}%
  \BibitemOpen
  \bibfield  {author} {\bibinfo {author} {\bibfnamefont {X.}~\bibnamefont
  {Yang}}, \bibinfo {author} {\bibfnamefont {K.}~\bibnamefont {Parrish}},
  \bibinfo {author} {\bibfnamefont {Y.-L.}\ \bibnamefont {Li}}, \bibinfo
  {author} {\bibfnamefont {B.}~\bibnamefont {Sa}}, \bibinfo {author}
  {\bibfnamefont {H.}~\bibnamefont {Zhan}}, \ and\ \bibinfo {author}
  {\bibfnamefont {Q.}~\bibnamefont {Zhu}},\ }\href {\doibase
  10.1103/PhysRevB.103.125103} {\bibfield  {journal} {\bibinfo  {journal}
  {Phys. Rev. B}\ }\textbf {\bibinfo {volume} {103}},\ \bibinfo {pages}
  {125103} (\bibinfo {year} {2021})}\BibitemShut {NoStop}%
\bibitem [{\citenamefont {Kang}\ \emph {et~al.}(2020)\citenamefont {Kang},
  \citenamefont {Bang}, \citenamefont {Chung}, \citenamefont {Nandadasa},
  \citenamefont {Han}, \citenamefont {Lee}, \citenamefont {Lee}, \citenamefont
  {Lee}, \citenamefont {Ma}, \citenamefont {Oh} \emph
  {et~al.}}]{kang2020water}%
  \BibitemOpen
  \bibfield  {author} {\bibinfo {author} {\bibfnamefont {S.~H.}\ \bibnamefont
  {Kang}}, \bibinfo {author} {\bibfnamefont {J.}~\bibnamefont {Bang}}, \bibinfo
  {author} {\bibfnamefont {K.}~\bibnamefont {Chung}}, \bibinfo {author}
  {\bibfnamefont {C.~N.}\ \bibnamefont {Nandadasa}}, \bibinfo {author}
  {\bibfnamefont {G.}~\bibnamefont {Han}}, \bibinfo {author} {\bibfnamefont
  {S.}~\bibnamefont {Lee}}, \bibinfo {author} {\bibfnamefont {K.~H.}\
  \bibnamefont {Lee}}, \bibinfo {author} {\bibfnamefont {K.}~\bibnamefont
  {Lee}}, \bibinfo {author} {\bibfnamefont {Y.}~\bibnamefont {Ma}}, \bibinfo
  {author} {\bibfnamefont {S.~H.}\ \bibnamefont {Oh}},  \emph {et~al.},\ }\href
  {\doibase 10.1126/sciadv.aba7416} {\bibfield  {journal} {\bibinfo  {journal}
  {Sci. Adv.}\ }\textbf {\bibinfo {volume} {6}},\ \bibinfo {pages} {eaba7416}
  (\bibinfo {year} {2020})}\BibitemShut {NoStop}%
\bibitem [{\citenamefont {Boysen}\ \emph {et~al.}(2007)\citenamefont {Boysen},
  \citenamefont {Lerch}, \citenamefont {Stys},\ and\ \citenamefont
  {Senyshyn}}]{boysen2007structure}%
  \BibitemOpen
  \bibfield  {author} {\bibinfo {author} {\bibfnamefont {H.}~\bibnamefont
  {Boysen}}, \bibinfo {author} {\bibfnamefont {M.}~\bibnamefont {Lerch}},
  \bibinfo {author} {\bibfnamefont {A.}~\bibnamefont {Stys}}, \ and\ \bibinfo
  {author} {\bibfnamefont {A.}~\bibnamefont {Senyshyn}},\ }\href {\doibase
  10.1107/S0108768107030005} {\bibfield  {journal} {\bibinfo  {journal} {Acta
  Cryst. B}\ }\textbf {\bibinfo {volume} {63}},\ \bibinfo {pages} {675}
  (\bibinfo {year} {2007})}\BibitemShut {NoStop}%
\bibitem [{\citenamefont {Miyakawa}\ \emph {et~al.}(2007)\citenamefont
  {Miyakawa}, \citenamefont {Kim}, \citenamefont {Hirano}, \citenamefont
  {Kohama}, \citenamefont {Kawaji}, \citenamefont {Atake}, \citenamefont
  {Ikegami}, \citenamefont {Kono},\ and\ \citenamefont
  {Hosono}}]{Miyakawa-JACS-2007}%
  \BibitemOpen
  \bibfield  {author} {\bibinfo {author} {\bibfnamefont {M.}~\bibnamefont
  {Miyakawa}}, \bibinfo {author} {\bibfnamefont {S.~W.}\ \bibnamefont {Kim}},
  \bibinfo {author} {\bibfnamefont {M.}~\bibnamefont {Hirano}}, \bibinfo
  {author} {\bibfnamefont {Y.}~\bibnamefont {Kohama}}, \bibinfo {author}
  {\bibfnamefont {H.}~\bibnamefont {Kawaji}}, \bibinfo {author} {\bibfnamefont
  {T.}~\bibnamefont {Atake}}, \bibinfo {author} {\bibfnamefont
  {H.}~\bibnamefont {Ikegami}}, \bibinfo {author} {\bibfnamefont
  {K.}~\bibnamefont {Kono}}, \ and\ \bibinfo {author} {\bibfnamefont
  {H.}~\bibnamefont {Hosono}},\ }\href {\doibase 10.1021/ja0724644} {\bibfield
  {journal} {\bibinfo  {journal} {J. Am. Chem. Soc.}\ }\textbf {\bibinfo
  {volume} {129}},\ \bibinfo {pages} {7270} (\bibinfo {year}
  {2007})}\BibitemShut {NoStop}%
\bibitem [{\citenamefont {Kitano}\ \emph {et~al.}(2012)\citenamefont {Kitano},
  \citenamefont {Inoue}, \citenamefont {Yamazaki}, \citenamefont {Hayashi},
  \citenamefont {Kanbara}, \citenamefont {Matsuishi}, \citenamefont {Yokoyama},
  \citenamefont {Kim}, \citenamefont {Hara},\ and\ \citenamefont
  {Hosono}}]{Kitano-NChem-2012}%
  \BibitemOpen
  \bibfield  {author} {\bibinfo {author} {\bibfnamefont {M.}~\bibnamefont
  {Kitano}}, \bibinfo {author} {\bibfnamefont {Y.}~\bibnamefont {Inoue}},
  \bibinfo {author} {\bibfnamefont {Y.}~\bibnamefont {Yamazaki}}, \bibinfo
  {author} {\bibfnamefont {F.}~\bibnamefont {Hayashi}}, \bibinfo {author}
  {\bibfnamefont {S.}~\bibnamefont {Kanbara}}, \bibinfo {author} {\bibfnamefont
  {S.}~\bibnamefont {Matsuishi}}, \bibinfo {author} {\bibfnamefont
  {T.}~\bibnamefont {Yokoyama}}, \bibinfo {author} {\bibfnamefont {S.-W.}\
  \bibnamefont {Kim}}, \bibinfo {author} {\bibfnamefont {M.}~\bibnamefont
  {Hara}}, \ and\ \bibinfo {author} {\bibfnamefont {H.}~\bibnamefont
  {Hosono}},\ }\href {\doibase 10.1038/nchem.1476} {\bibfield  {journal}
  {\bibinfo  {journal} {Nat. Chem.}\ }\textbf {\bibinfo {volume} {4}},\
  \bibinfo {pages} {934} (\bibinfo {year} {2012})}\BibitemShut {NoStop}%
\bibitem [{\citenamefont {Kuganathan}\ \emph {et~al.}(2014)\citenamefont
  {Kuganathan}, \citenamefont {Hosono}, \citenamefont {Shluger},\ and\
  \citenamefont {Sushko}}]{Kuganathan-JACS-2014}%
  \BibitemOpen
  \bibfield  {author} {\bibinfo {author} {\bibfnamefont {N.}~\bibnamefont
  {Kuganathan}}, \bibinfo {author} {\bibfnamefont {H.}~\bibnamefont {Hosono}},
  \bibinfo {author} {\bibfnamefont {A.~L.}\ \bibnamefont {Shluger}}, \ and\
  \bibinfo {author} {\bibfnamefont {P.~V.}\ \bibnamefont {Sushko}},\ }\href
  {\doibase 10.1021/ja410925g} {\bibfield  {journal} {\bibinfo  {journal} {J.
  Am. Chem. Soc.}\ }\textbf {\bibinfo {volume} {136}},\ \bibinfo {pages} {2216}
  (\bibinfo {year} {2014})}\BibitemShut {NoStop}%
\bibitem [{\citenamefont {Hayashi}\ \emph {et~al.}(2014)\citenamefont
  {Hayashi}, \citenamefont {Tomota}, \citenamefont {Kitano}, \citenamefont
  {Toda}, \citenamefont {Yokoyama},\ and\ \citenamefont
  {Hosono}}]{Hayashi-JACS-2014}%
  \BibitemOpen
  \bibfield  {author} {\bibinfo {author} {\bibfnamefont {F.}~\bibnamefont
  {Hayashi}}, \bibinfo {author} {\bibfnamefont {Y.}~\bibnamefont {Tomota}},
  \bibinfo {author} {\bibfnamefont {M.}~\bibnamefont {Kitano}}, \bibinfo
  {author} {\bibfnamefont {Y.}~\bibnamefont {Toda}}, \bibinfo {author}
  {\bibfnamefont {T.}~\bibnamefont {Yokoyama}}, \ and\ \bibinfo {author}
  {\bibfnamefont {H.}~\bibnamefont {Hosono}},\ }\href {\doibase
  10.1021/ja504185m} {\bibfield  {journal} {\bibinfo  {journal} {J. Am. Chem.
  Soc.}\ }\textbf {\bibinfo {volume} {136}},\ \bibinfo {pages} {11698}
  (\bibinfo {year} {2014})}\BibitemShut {NoStop}%
\bibitem [{\citenamefont {Hosono}\ \emph {et~al.}(2017)\citenamefont {Hosono},
  \citenamefont {Kim}, \citenamefont {Toda}, \citenamefont {Kamiya},\ and\
  \citenamefont {Watanabe}}]{Hosono-PNAS-2017}%
  \BibitemOpen
  \bibfield  {author} {\bibinfo {author} {\bibfnamefont {H.}~\bibnamefont
  {Hosono}}, \bibinfo {author} {\bibfnamefont {J.}~\bibnamefont {Kim}},
  \bibinfo {author} {\bibfnamefont {Y.}~\bibnamefont {Toda}}, \bibinfo {author}
  {\bibfnamefont {T.}~\bibnamefont {Kamiya}}, \ and\ \bibinfo {author}
  {\bibfnamefont {S.}~\bibnamefont {Watanabe}},\ }\href {\doibase
  10.1073/pnas.1617186114} {\bibfield  {journal} {\bibinfo  {journal} {Proc.
  Natl. Acad. Sci.}\ }\textbf {\bibinfo {volume} {114}},\ \bibinfo {pages}
  {233} (\bibinfo {year} {2017})}\BibitemShut {NoStop}%
\bibitem [{\citenamefont {Antao}\ and\ \citenamefont
  {Hassan}(2002)}]{antao2002thermal}%
  \BibitemOpen
  \bibfield  {author} {\bibinfo {author} {\bibfnamefont {S.~M.}\ \bibnamefont
  {Antao}}\ and\ \bibinfo {author} {\bibfnamefont {I.}~\bibnamefont {Hassan}},\
  }\href {\doibase 10.2113/gscanmin.40.1.163} {\bibfield  {journal} {\bibinfo
  {journal} {Can. Mineral.}\ }\textbf {\bibinfo {volume} {40}},\ \bibinfo
  {pages} {163} (\bibinfo {year} {2002})}\BibitemShut {NoStop}%
\bibitem [{\citenamefont {Kang}\ and\ \citenamefont
  {Estreicher}(2014)}]{kang_prb2014}%
  \BibitemOpen
  \bibfield  {author} {\bibinfo {author} {\bibfnamefont {B.}~\bibnamefont
  {Kang}}\ and\ \bibinfo {author} {\bibfnamefont {S.}~\bibnamefont
  {Estreicher}},\ }\href {\doibase 10.1103/PhysRevB.89.155409} {\bibfield
  {journal} {\bibinfo  {journal} {Phys. Rev. B}\ }\textbf {\bibinfo {volume}
  {89}},\ \bibinfo {pages} {155409} (\bibinfo {year} {2014})}\BibitemShut
  {NoStop}%
\bibitem [{\citenamefont {Bl\"ochl}(1994)}]{PAW-PRB-1994}%
  \BibitemOpen
  \bibfield  {author} {\bibinfo {author} {\bibfnamefont {P.~E.}\ \bibnamefont
  {Bl\"ochl}},\ }\href {\doibase 10.1103/PhysRevB.50.17953} {\bibfield
  {journal} {\bibinfo  {journal} {Phys. Rev. B}\ }\textbf {\bibinfo {volume}
  {50}},\ \bibinfo {pages} {17953} (\bibinfo {year} {1994})}\BibitemShut
  {NoStop}%
\bibitem [{\citenamefont {Kresse}\ and\ \citenamefont
  {Furthm\"uller}(1996)}]{Vasp-PRB-1996}%
  \BibitemOpen
  \bibfield  {author} {\bibinfo {author} {\bibfnamefont {G.}~\bibnamefont
  {Kresse}}\ and\ \bibinfo {author} {\bibfnamefont {J.}~\bibnamefont
  {Furthm\"uller}},\ }\href {\doibase 10.1103/PhysRevB.54.11169} {\bibfield
  {journal} {\bibinfo  {journal} {Phys. Rev. B}\ }\textbf {\bibinfo {volume}
  {54}},\ \bibinfo {pages} {11169} (\bibinfo {year} {1996})}\BibitemShut
  {NoStop}%
\bibitem [{\citenamefont {Kresse}\ and\ \citenamefont
  {Joubert}(1999)}]{kresse1999ultrasoft}%
  \BibitemOpen
  \bibfield  {author} {\bibinfo {author} {\bibfnamefont {G.}~\bibnamefont
  {Kresse}}\ and\ \bibinfo {author} {\bibfnamefont {D.}~\bibnamefont
  {Joubert}},\ }\href {\doibase 10.1103/PhysRevB.59.1758} {\bibfield  {journal}
  {\bibinfo  {journal} {Phys. Rev. B}\ }\textbf {\bibinfo {volume} {59}},\
  \bibinfo {pages} {1758} (\bibinfo {year} {1999})}\BibitemShut {NoStop}%
\bibitem [{\citenamefont {Perdew}\ \emph {et~al.}(1996)\citenamefont {Perdew},
  \citenamefont {Burke},\ and\ \citenamefont {Ernzerhof}}]{PBE-PRL-1996}%
  \BibitemOpen
  \bibfield  {author} {\bibinfo {author} {\bibfnamefont {J.~P.}\ \bibnamefont
  {Perdew}}, \bibinfo {author} {\bibfnamefont {K.}~\bibnamefont {Burke}}, \
  and\ \bibinfo {author} {\bibfnamefont {M.}~\bibnamefont {Ernzerhof}},\ }\href
  {\doibase 10.1103/PhysRevLett.77.3865} {\bibfield  {journal} {\bibinfo
  {journal} {Phys. Rev. Lett.}\ }\textbf {\bibinfo {volume} {77}},\ \bibinfo
  {pages} {3865} (\bibinfo {year} {1996})}\BibitemShut {NoStop}%
\bibitem [{\citenamefont {Togo}\ \emph {et~al.}(2008)\citenamefont {Togo},
  \citenamefont {Oba},\ and\ \citenamefont {Tanaka}}]{Togo-PRB-2008}%
  \BibitemOpen
  \bibfield  {author} {\bibinfo {author} {\bibfnamefont {A.}~\bibnamefont
  {Togo}}, \bibinfo {author} {\bibfnamefont {F.}~\bibnamefont {Oba}}, \ and\
  \bibinfo {author} {\bibfnamefont {I.}~\bibnamefont {Tanaka}},\ }\href
  {\doibase 10.1103/PhysRevB.78.134106} {\bibfield  {journal} {\bibinfo
  {journal} {Phys. Rev. B}\ }\textbf {\bibinfo {volume} {78}},\ \bibinfo
  {pages} {134106} (\bibinfo {year} {2008})}\BibitemShut {NoStop}%
\bibitem [{\citenamefont {Pauling}(1930)}]{linux_zkri1930}%
  \BibitemOpen
  \bibfield  {author} {\bibinfo {author} {\bibfnamefont {L.}~\bibnamefont
  {Pauling}},\ }\href {\doibase 10.1524/zkri.1930.74.1.213} {\bibfield
  {journal} {\bibinfo  {journal} {Z. Kristallogr}\ }\textbf {\bibinfo {volume}
  {74}},\ \bibinfo {pages} {213} (\bibinfo {year} {1930})}\BibitemShut
  {NoStop}%
\bibitem [{\citenamefont {McMullan}\ \emph {et~al.}(1996)\citenamefont
  {McMullan}, \citenamefont {Ghose}, \citenamefont {Haga},\ and\ \citenamefont
  {Schomaker}}]{mcmullan_actacryst1996}%
  \BibitemOpen
  \bibfield  {author} {\bibinfo {author} {\bibfnamefont {R.}~\bibnamefont
  {McMullan}}, \bibinfo {author} {\bibfnamefont {S.}~\bibnamefont {Ghose}},
  \bibinfo {author} {\bibfnamefont {N.}~\bibnamefont {Haga}}, \ and\ \bibinfo
  {author} {\bibfnamefont {V.}~\bibnamefont {Schomaker}},\ }\href {\doibase
  10.1107/S0108768196004132} {\bibfield  {journal} {\bibinfo  {journal} {Acta
  Cryst. B}\ }\textbf {\bibinfo {volume} {52}},\ \bibinfo {pages} {616}
  (\bibinfo {year} {1996})}\BibitemShut {NoStop}%
\bibitem [{\citenamefont {Hassan}\ \emph {et~al.}(2004)\citenamefont {Hassan},
  \citenamefont {Antao},\ and\ \citenamefont {Parise}}]{ishmael_am2004}%
  \BibitemOpen
  \bibfield  {author} {\bibinfo {author} {\bibfnamefont {I.}~\bibnamefont
  {Hassan}}, \bibinfo {author} {\bibfnamefont {S.~M.}\ \bibnamefont {Antao}}, \
  and\ \bibinfo {author} {\bibfnamefont {J.~B.}\ \bibnamefont {Parise}},\
  }\href {\doibase 10.2138/am-2004-2-315} {\bibfield  {journal} {\bibinfo
  {journal} {Am. Mineral.}\ }\textbf {\bibinfo {volume} {89}},\ \bibinfo
  {pages} {359} (\bibinfo {year} {2004})}\BibitemShut {NoStop}%
\bibitem [{\citenamefont {Nielsen}\ \emph {et~al.}(1991)\citenamefont
  {Nielsen}, \citenamefont {Bilds{\o}e}, \citenamefont {Jakobsen},\ and\
  \citenamefont {Norby}}]{niels_zeolites1991}%
  \BibitemOpen
  \bibfield  {author} {\bibinfo {author} {\bibfnamefont {N.~C.}\ \bibnamefont
  {Nielsen}}, \bibinfo {author} {\bibfnamefont {H.}~\bibnamefont {Bilds{\o}e}},
  \bibinfo {author} {\bibfnamefont {H.~J.}\ \bibnamefont {Jakobsen}}, \ and\
  \bibinfo {author} {\bibfnamefont {P.}~\bibnamefont {Norby}},\ }\href
  {\doibase 10.1016/S0144-2449(05)80015-6} {\bibfield  {journal} {\bibinfo
  {journal} {Zeolites}\ }\textbf {\bibinfo {volume} {11}},\ \bibinfo {pages}
  {622} (\bibinfo {year} {1991})}\BibitemShut {NoStop}%
\bibitem [{\citenamefont {Van~Doorn}\ \emph {et~al.}(1972)\citenamefont
  {Van~Doorn}, \citenamefont {Schipper},\ and\ \citenamefont
  {Bolwijn}}]{doorn_jes1972}%
  \BibitemOpen
  \bibfield  {author} {\bibinfo {author} {\bibfnamefont {C.}~\bibnamefont
  {Van~Doorn}}, \bibinfo {author} {\bibfnamefont {D.}~\bibnamefont {Schipper}},
  \ and\ \bibinfo {author} {\bibfnamefont {P.}~\bibnamefont {Bolwijn}},\ }\href
  {\doibase 10.1149/1.2404141} {\bibfield  {journal} {\bibinfo  {journal} {J.
  Electrochem. Soc.}\ }\textbf {\bibinfo {volume} {119}},\ \bibinfo {pages}
  {85} (\bibinfo {year} {1972})}\BibitemShut {NoStop}%
\bibitem [{\citenamefont {Johnson}\ and\ \citenamefont
  {Weller}(1997)}]{geoffrey_pzmm1997}%
  \BibitemOpen
  \bibfield  {author} {\bibinfo {author} {\bibfnamefont {G.~M.}\ \bibnamefont
  {Johnson}}\ and\ \bibinfo {author} {\bibfnamefont {M.~T.}\ \bibnamefont
  {Weller}},\ }in\ \href {\doibase 10.1016/S0167-2991(97)80565-4} {\emph
  {\bibinfo {booktitle} {Studies in Surface Science and Catalysis}}},\ Vol.\
  \bibinfo {volume} {105}\ (\bibinfo  {publisher} {Elsevier},\ \bibinfo {year}
  {1997})\ pp.\ \bibinfo {pages} {269--275}\BibitemShut {NoStop}%
\bibitem [{\citenamefont {Grimme}\ \emph {et~al.}(2010)\citenamefont {Grimme},
  \citenamefont {Antony}, \citenamefont {Ehrlich},\ and\ \citenamefont
  {Krieg}}]{stefan_jcp2010}%
  \BibitemOpen
  \bibfield  {author} {\bibinfo {author} {\bibfnamefont {S.}~\bibnamefont
  {Grimme}}, \bibinfo {author} {\bibfnamefont {J.}~\bibnamefont {Antony}},
  \bibinfo {author} {\bibfnamefont {S.}~\bibnamefont {Ehrlich}}, \ and\
  \bibinfo {author} {\bibfnamefont {H.}~\bibnamefont {Krieg}},\ }\href
  {\doibase 10.1063/1.3382344} {\bibfield  {journal} {\bibinfo  {journal} {J.
  Chem. Phys.}\ }\textbf {\bibinfo {volume} {132}},\ \bibinfo {pages} {154104}
  (\bibinfo {year} {2010})}\BibitemShut {NoStop}%
\bibitem [{\citenamefont {Mofrad}\ \emph {et~al.}(2018)\citenamefont {Mofrad},
  \citenamefont {Peixoto}, \citenamefont {Blumeyer}, \citenamefont {Liu},
  \citenamefont {Hunt},\ and\ \citenamefont {Hammond}}]{amir_jpcc2018}%
  \BibitemOpen
  \bibfield  {author} {\bibinfo {author} {\bibfnamefont {A.~M.}\ \bibnamefont
  {Mofrad}}, \bibinfo {author} {\bibfnamefont {C.}~\bibnamefont {Peixoto}},
  \bibinfo {author} {\bibfnamefont {J.}~\bibnamefont {Blumeyer}}, \bibinfo
  {author} {\bibfnamefont {J.}~\bibnamefont {Liu}}, \bibinfo {author}
  {\bibfnamefont {H.~K.}\ \bibnamefont {Hunt}}, \ and\ \bibinfo {author}
  {\bibfnamefont {K.~D.}\ \bibnamefont {Hammond}},\ }\href {\doibase
  10.1021/acs.jpcc.8b07633} {\bibfield  {journal} {\bibinfo  {journal} {J.
  Phys. Chem. C}\ }\textbf {\bibinfo {volume} {122}},\ \bibinfo {pages} {24765}
  (\bibinfo {year} {2018})}\BibitemShut {NoStop}%
\bibitem [{\citenamefont {Thomson}\ and\ \citenamefont
  {Wentzcovitch}(1998)}]{kendall_jcp1998}%
  \BibitemOpen
  \bibfield  {author} {\bibinfo {author} {\bibfnamefont {K.~T.}\ \bibnamefont
  {Thomson}}\ and\ \bibinfo {author} {\bibfnamefont {R.~M.}\ \bibnamefont
  {Wentzcovitch}},\ }\href {\doibase 10.1063/1.476287} {\bibfield  {journal}
  {\bibinfo  {journal} {J. Chem. Phys.}\ }\textbf {\bibinfo {volume} {108}},\
  \bibinfo {pages} {8584} (\bibinfo {year} {1998})}\BibitemShut {NoStop}%
\bibitem [{\citenamefont {Ulian}\ and\ \citenamefont
  {Valdr{\`e}}(2022)}]{ulian_mdpi2022}%
  \BibitemOpen
  \bibfield  {author} {\bibinfo {author} {\bibfnamefont {G.}~\bibnamefont
  {Ulian}}\ and\ \bibinfo {author} {\bibfnamefont {G.}~\bibnamefont
  {Valdr{\`e}}},\ }\href {\doibase 10.3390/min12101323} {\bibfield  {journal}
  {\bibinfo  {journal} {Minerals}\ }\textbf {\bibinfo {volume} {12}},\ \bibinfo
  {pages} {1323} (\bibinfo {year} {2022})}\BibitemShut {NoStop}%
\bibitem [{\citenamefont {Pan}\ \emph {et~al.}(2016)\citenamefont {Pan},
  \citenamefont {Liu}, \citenamefont {Chen}, \citenamefont {Yan}, \citenamefont
  {Yang},\ and\ \citenamefont {Yu}}]{lijun_jm2016}%
  \BibitemOpen
  \bibfield  {author} {\bibinfo {author} {\bibfnamefont {L.}~\bibnamefont
  {Pan}}, \bibinfo {author} {\bibfnamefont {W.}~\bibnamefont {Liu}}, \bibinfo
  {author} {\bibfnamefont {W.}~\bibnamefont {Chen}}, \bibinfo {author}
  {\bibfnamefont {K.}~\bibnamefont {Yan}}, \bibinfo {author} {\bibfnamefont
  {H.}~\bibnamefont {Yang}}, \ and\ \bibinfo {author} {\bibfnamefont
  {J.}~\bibnamefont {Yu}},\ }\href {\doibase 10.1016/j.molstruc.2015.10.045}
  {\bibfield  {journal} {\bibinfo  {journal} {J. Mol. Struct.}\ }\textbf
  {\bibinfo {volume} {1106}},\ \bibinfo {pages} {59} (\bibinfo {year}
  {2016})}\BibitemShut {NoStop}%
\bibitem [{\citenamefont {Filippone}\ \emph {et~al.}(1995)\citenamefont
  {Filippone}, \citenamefont {Buda}, \citenamefont {Iarlori}, \citenamefont
  {Moretti},\ and\ \citenamefont {Porta}}]{francesco_jpc1995}%
  \BibitemOpen
  \bibfield  {author} {\bibinfo {author} {\bibfnamefont {F.}~\bibnamefont
  {Filippone}}, \bibinfo {author} {\bibfnamefont {F.}~\bibnamefont {Buda}},
  \bibinfo {author} {\bibfnamefont {S.}~\bibnamefont {Iarlori}}, \bibinfo
  {author} {\bibfnamefont {G.}~\bibnamefont {Moretti}}, \ and\ \bibinfo
  {author} {\bibfnamefont {P.}~\bibnamefont {Porta}},\ }\href {\doibase
  10.1021/j100034a029} {\bibfield  {journal} {\bibinfo  {journal} {J. Phys.
  Chem.}\ }\textbf {\bibinfo {volume} {99}},\ \bibinfo {pages} {12883}
  (\bibinfo {year} {1995})}\BibitemShut {NoStop}%
\bibitem [{\citenamefont {Cano}\ \emph {et~al.}(2010)\citenamefont {Cano},
  \citenamefont {Ayta},\ and\ \citenamefont {Watanabe}}]{nilo_ssc2010}%
  \BibitemOpen
  \bibfield  {author} {\bibinfo {author} {\bibfnamefont {N.~F.}\ \bibnamefont
  {Cano}}, \bibinfo {author} {\bibfnamefont {W.~E.}\ \bibnamefont {Ayta}}, \
  and\ \bibinfo {author} {\bibfnamefont {S.}~\bibnamefont {Watanabe}},\ }\href
  {\doibase 10.1016/j.ssc.2009.10.034} {\bibfield  {journal} {\bibinfo
  {journal} {Solid State Commun.}\ }\textbf {\bibinfo {volume} {150}},\
  \bibinfo {pages} {195} (\bibinfo {year} {2010})}\BibitemShut {NoStop}%
\bibitem [{\citenamefont {Hosono}(2013)}]{hosono2013exploring}%
  \BibitemOpen
  \bibfield  {author} {\bibinfo {author} {\bibfnamefont {H.}~\bibnamefont
  {Hosono}},\ }\href {\doibase 10.7567/JJAP.52.090001} {\bibfield  {journal}
  {\bibinfo  {journal} {Jpn. J. Appl. Phys.}\ }\textbf {\bibinfo {volume}
  {52}},\ \bibinfo {pages} {090001} (\bibinfo {year} {2013})}\BibitemShut
  {NoStop}%
\bibitem [{\citenamefont {Srdanov}\ \emph {et~al.}(1998)\citenamefont
  {Srdanov}, \citenamefont {Stucky}, \citenamefont {Lippmaa},\ and\
  \citenamefont {Engelhardt}}]{Srdanov-PRL-1998}%
  \BibitemOpen
  \bibfield  {author} {\bibinfo {author} {\bibfnamefont {V.~I.}\ \bibnamefont
  {Srdanov}}, \bibinfo {author} {\bibfnamefont {G.~D.}\ \bibnamefont {Stucky}},
  \bibinfo {author} {\bibfnamefont {E.}~\bibnamefont {Lippmaa}}, \ and\
  \bibinfo {author} {\bibfnamefont {G.}~\bibnamefont {Engelhardt}},\ }\href
  {\doibase 10.1103/PhysRevLett.80.2449} {\bibfield  {journal} {\bibinfo
  {journal} {Phys. Rev. Lett.}\ }\textbf {\bibinfo {volume} {80}},\ \bibinfo
  {pages} {2449} (\bibinfo {year} {1998})}\BibitemShut {NoStop}%
\bibitem [{\citenamefont {Madsen}\ \emph {et~al.}(2001)\citenamefont {Madsen},
  \citenamefont {Iversen}, \citenamefont {Blaha},\ and\ \citenamefont
  {Schwarz}}]{madsen2001electronic}%
  \BibitemOpen
  \bibfield  {author} {\bibinfo {author} {\bibfnamefont {G.~K.}\ \bibnamefont
  {Madsen}}, \bibinfo {author} {\bibfnamefont {B.~B.}\ \bibnamefont {Iversen}},
  \bibinfo {author} {\bibfnamefont {P.}~\bibnamefont {Blaha}}, \ and\ \bibinfo
  {author} {\bibfnamefont {K.}~\bibnamefont {Schwarz}},\ }\href {\doibase
  10.1103/PhysRevB.64.195102} {\bibfield  {journal} {\bibinfo  {journal} {Phys.
  Rev. B}\ }\textbf {\bibinfo {volume} {64}},\ \bibinfo {pages} {195102}
  (\bibinfo {year} {2001})}\BibitemShut {NoStop}%
\bibitem [{\citenamefont {Srdanov}\ \emph {et~al.}(1992)\citenamefont
  {Srdanov}, \citenamefont {Haug}, \citenamefont {Metiu},\ and\ \citenamefont
  {Stucky}}]{srdanov_jpc1992}%
  \BibitemOpen
  \bibfield  {author} {\bibinfo {author} {\bibfnamefont {V.}~\bibnamefont
  {Srdanov}}, \bibinfo {author} {\bibfnamefont {K.}~\bibnamefont {Haug}},
  \bibinfo {author} {\bibfnamefont {H.}~\bibnamefont {Metiu}}, \ and\ \bibinfo
  {author} {\bibfnamefont {G.}~\bibnamefont {Stucky}},\ }\href {\doibase
  10.1021/j100201a064} {\bibfield  {journal} {\bibinfo  {journal} {J. Phys.
  Chem.}\ }\textbf {\bibinfo {volume} {96}},\ \bibinfo {pages} {9039} (\bibinfo
  {year} {1992})}\BibitemShut {NoStop}%
\bibitem [{\citenamefont {Smeulders}\ \emph {et~al.}(1987)\citenamefont
  {Smeulders}, \citenamefont {Hefni}, \citenamefont {Klaassen}, \citenamefont
  {De~Boer}, \citenamefont {Westphal},\ and\ \citenamefont
  {Geismar}}]{smeulders_zeolites1987}%
  \BibitemOpen
  \bibfield  {author} {\bibinfo {author} {\bibfnamefont {J.}~\bibnamefont
  {Smeulders}}, \bibinfo {author} {\bibfnamefont {M.}~\bibnamefont {Hefni}},
  \bibinfo {author} {\bibfnamefont {A.}~\bibnamefont {Klaassen}}, \bibinfo
  {author} {\bibfnamefont {E.}~\bibnamefont {De~Boer}}, \bibinfo {author}
  {\bibfnamefont {U.}~\bibnamefont {Westphal}}, \ and\ \bibinfo {author}
  {\bibfnamefont {G.}~\bibnamefont {Geismar}},\ }\href {\doibase
  10.1016/0144-2449(87)90038-8} {\bibfield  {journal} {\bibinfo  {journal}
  {Zeolites}\ }\textbf {\bibinfo {volume} {7}},\ \bibinfo {pages} {347}
  (\bibinfo {year} {1987})}\BibitemShut {NoStop}%
\bibitem [{\citenamefont {Blake}\ \emph {et~al.}(1996)\citenamefont {Blake},
  \citenamefont {Srdanov}, \citenamefont {Stucky},\ and\ \citenamefont
  {Metiu}}]{nick_jcp1996}%
  \BibitemOpen
  \bibfield  {author} {\bibinfo {author} {\bibfnamefont {N.~P.}\ \bibnamefont
  {Blake}}, \bibinfo {author} {\bibfnamefont {V.~I.}\ \bibnamefont {Srdanov}},
  \bibinfo {author} {\bibfnamefont {G.~D.}\ \bibnamefont {Stucky}}, \ and\
  \bibinfo {author} {\bibfnamefont {H.}~\bibnamefont {Metiu}},\ }\href
  {\doibase 10.1063/1.471561} {\bibfield  {journal} {\bibinfo  {journal} {J.
  Chem. Phys.}\ }\textbf {\bibinfo {volume} {104}},\ \bibinfo {pages} {8721}
  (\bibinfo {year} {1996})}\BibitemShut {NoStop}%
\bibitem [{\citenamefont {Stoliaroff}\ \emph {et~al.}(2021)\citenamefont
  {Stoliaroff}, \citenamefont {Schira}, \citenamefont {Blumentritt},
  \citenamefont {Fritsch}, \citenamefont {Jobic},\ and\ \citenamefont
  {Latouche}}]{adreien_jpcc2021}%
  \BibitemOpen
  \bibfield  {author} {\bibinfo {author} {\bibfnamefont {A.}~\bibnamefont
  {Stoliaroff}}, \bibinfo {author} {\bibfnamefont {R.}~\bibnamefont {Schira}},
  \bibinfo {author} {\bibfnamefont {F.}~\bibnamefont {Blumentritt}}, \bibinfo
  {author} {\bibfnamefont {E.}~\bibnamefont {Fritsch}}, \bibinfo {author}
  {\bibfnamefont {S.}~\bibnamefont {Jobic}}, \ and\ \bibinfo {author}
  {\bibfnamefont {C.}~\bibnamefont {Latouche}},\ }\href {\doibase
  10.1021/acs.jpcc.1c02423} {\bibfield  {journal} {\bibinfo  {journal} {J.
  Phys. Chem. C}\ }\textbf {\bibinfo {volume} {125}},\ \bibinfo {pages} {16674}
  (\bibinfo {year} {2021})}\BibitemShut {NoStop}%
\bibitem [{\citenamefont {Colinet}\ \emph {et~al.}(2020)\citenamefont
  {Colinet}, \citenamefont {Gheeraert}, \citenamefont {Curutchet},\ and\
  \citenamefont {Le~Bahers}}]{pauline_jpcc2020}%
  \BibitemOpen
  \bibfield  {author} {\bibinfo {author} {\bibfnamefont {P.}~\bibnamefont
  {Colinet}}, \bibinfo {author} {\bibfnamefont {A.}~\bibnamefont {Gheeraert}},
  \bibinfo {author} {\bibfnamefont {A.}~\bibnamefont {Curutchet}}, \ and\
  \bibinfo {author} {\bibfnamefont {T.}~\bibnamefont {Le~Bahers}},\ }\href
  {\doibase 10.1021/acs.jpcc.0c00615} {\bibfield  {journal} {\bibinfo
  {journal} {J. Phys. Chem. C}\ }\textbf {\bibinfo {volume} {124}},\ \bibinfo
  {pages} {8949} (\bibinfo {year} {2020})}\BibitemShut {NoStop}%
\bibitem [{\citenamefont {Holton}\ and\ \citenamefont
  {Blum}(1962)}]{holton_prb1962}%
  \BibitemOpen
  \bibfield  {author} {\bibinfo {author} {\bibfnamefont {W.~C.}\ \bibnamefont
  {Holton}}\ and\ \bibinfo {author} {\bibfnamefont {H.}~\bibnamefont {Blum}},\
  }\href {\doibase 10.1103/PhysRev.125.89} {\bibfield  {journal} {\bibinfo
  {journal} {Phys. Rev.}\ }\textbf {\bibinfo {volume} {125}},\ \bibinfo {pages}
  {89} (\bibinfo {year} {1962})}\BibitemShut {NoStop}%
\bibitem [{\citenamefont {Koyama}\ and\ \citenamefont
  {Suemoto}(2011)}]{takeshi_rpp2011}%
  \BibitemOpen
  \bibfield  {author} {\bibinfo {author} {\bibfnamefont {T.}~\bibnamefont
  {Koyama}}\ and\ \bibinfo {author} {\bibfnamefont {T.}~\bibnamefont
  {Suemoto}},\ }\href {\doibase 10.1088/0034-4885/74/7/076502} {\bibfield
  {journal} {\bibinfo  {journal} {Rep. Prog. Phys.}\ }\textbf {\bibinfo
  {volume} {74}},\ \bibinfo {pages} {076502} (\bibinfo {year}
  {2011})}\BibitemShut {NoStop}%
\bibitem [{\citenamefont {McRae}\ \emph {et~al.}(2022)\citenamefont {McRae},
  \citenamefont {Radomsky}, \citenamefont {Pawlik}, \citenamefont {Druffel},
  \citenamefont {Sundberg}, \citenamefont {Lanetti}, \citenamefont {Donley},
  \citenamefont {White},\ and\ \citenamefont {Warren}}]{lauren_jacs2022}%
  \BibitemOpen
  \bibfield  {author} {\bibinfo {author} {\bibfnamefont {L.~M.}\ \bibnamefont
  {McRae}}, \bibinfo {author} {\bibfnamefont {R.~C.}\ \bibnamefont {Radomsky}},
  \bibinfo {author} {\bibfnamefont {J.~T.}\ \bibnamefont {Pawlik}}, \bibinfo
  {author} {\bibfnamefont {D.~L.}\ \bibnamefont {Druffel}}, \bibinfo {author}
  {\bibfnamefont {J.~D.}\ \bibnamefont {Sundberg}}, \bibinfo {author}
  {\bibfnamefont {M.~G.}\ \bibnamefont {Lanetti}}, \bibinfo {author}
  {\bibfnamefont {C.~L.}\ \bibnamefont {Donley}}, \bibinfo {author}
  {\bibfnamefont {K.~L.}\ \bibnamefont {White}}, \ and\ \bibinfo {author}
  {\bibfnamefont {S.~C.}\ \bibnamefont {Warren}},\ }\href {\doibase
  10.1021/jacs.2c03024} {\bibfield  {journal} {\bibinfo  {journal} {J. Am.
  Chem. Soc.}\ }\textbf {\bibinfo {volume} {144}},\ \bibinfo {pages}
  {10862–10869} (\bibinfo {year} {2022})}\BibitemShut {NoStop}%
\bibitem [{\citenamefont {Sankey}\ \emph {et~al.}(1998)\citenamefont {Sankey},
  \citenamefont {Demkov},\ and\ \citenamefont {Lenosky}}]{ott_prb1998}%
  \BibitemOpen
  \bibfield  {author} {\bibinfo {author} {\bibfnamefont {O.~F.}\ \bibnamefont
  {Sankey}}, \bibinfo {author} {\bibfnamefont {A.~A.}\ \bibnamefont {Demkov}},
  \ and\ \bibinfo {author} {\bibfnamefont {T.}~\bibnamefont {Lenosky}},\ }\href
  {\doibase 10.1103/PhysRevB.57.15129} {\bibfield  {journal} {\bibinfo
  {journal} {Phys. Rev. B}\ }\textbf {\bibinfo {volume} {57}},\ \bibinfo
  {pages} {15129} (\bibinfo {year} {1998})}\BibitemShut {NoStop}%
\bibitem [{\citenamefont {Heinmaa}\ \emph {et~al.}(2000)\citenamefont
  {Heinmaa}, \citenamefont {Vija},\ and\ \citenamefont
  {Lippmaa}}]{heinmaa_cpl2000}%
  \BibitemOpen
  \bibfield  {author} {\bibinfo {author} {\bibfnamefont {I.}~\bibnamefont
  {Heinmaa}}, \bibinfo {author} {\bibfnamefont {S.}~\bibnamefont {Vija}}, \
  and\ \bibinfo {author} {\bibfnamefont {E.}~\bibnamefont {Lippmaa}},\ }\href
  {\doibase 10.1016/S0009-2614(00)00859-9} {\bibfield  {journal} {\bibinfo
  {journal} {Chem. Phys. Lett.}\ }\textbf {\bibinfo {volume} {327}},\ \bibinfo
  {pages} {131} (\bibinfo {year} {2000})}\BibitemShut {NoStop}%
\bibitem [{\citenamefont {Becke}\ and\ \citenamefont {Edgecombe}(1990)}]{ELF}%
  \BibitemOpen
  \bibfield  {author} {\bibinfo {author} {\bibfnamefont {A.~D.}\ \bibnamefont
  {Becke}}\ and\ \bibinfo {author} {\bibfnamefont {K.~E.}\ \bibnamefont
  {Edgecombe}},\ }\href {\doibase 10.1063/1.458517} {\bibfield  {journal}
  {\bibinfo  {journal} {J. Chem. Phys.}\ }\textbf {\bibinfo {volume} {92}},\
  \bibinfo {pages} {5397} (\bibinfo {year} {1990})}\BibitemShut {NoStop}%
\bibitem [{\citenamefont {Petkov}\ \emph {et~al.}(2002)\citenamefont {Petkov},
  \citenamefont {Billinge}, \citenamefont {Vogt}, \citenamefont {Ichimura},\
  and\ \citenamefont {Dye}}]{Petkov-PRL-2002}%
  \BibitemOpen
  \bibfield  {author} {\bibinfo {author} {\bibfnamefont {V.}~\bibnamefont
  {Petkov}}, \bibinfo {author} {\bibfnamefont {S.~J.~L.}\ \bibnamefont
  {Billinge}}, \bibinfo {author} {\bibfnamefont {T.}~\bibnamefont {Vogt}},
  \bibinfo {author} {\bibfnamefont {A.~S.}\ \bibnamefont {Ichimura}}, \ and\
  \bibinfo {author} {\bibfnamefont {J.~L.}\ \bibnamefont {Dye}},\ }\href
  {\doibase 10.1103/PhysRevLett.89.075502} {\bibfield  {journal} {\bibinfo
  {journal} {Phys. Rev. Lett.}\ }\textbf {\bibinfo {volume} {89}},\ \bibinfo
  {pages} {075502} (\bibinfo {year} {2002})}\BibitemShut {NoStop}%
\bibitem [{\citenamefont {Wernette}\ \emph {et~al.}(2003)\citenamefont
  {Wernette}, \citenamefont {Ichimura}, \citenamefont {Urbin},\ and\
  \citenamefont {Dye}}]{wernette2003inorganic}%
  \BibitemOpen
  \bibfield  {author} {\bibinfo {author} {\bibfnamefont {D.~P.}\ \bibnamefont
  {Wernette}}, \bibinfo {author} {\bibfnamefont {A.~S.}\ \bibnamefont
  {Ichimura}}, \bibinfo {author} {\bibfnamefont {S.~A.}\ \bibnamefont {Urbin}},
  \ and\ \bibinfo {author} {\bibfnamefont {J.~L.}\ \bibnamefont {Dye}},\ }\href
  {\doibase 10.1021/cm020906z} {\bibfield  {journal} {\bibinfo  {journal}
  {Chem. Mater.}\ }\textbf {\bibinfo {volume} {15}},\ \bibinfo {pages} {1441}
  (\bibinfo {year} {2003})}\BibitemShut {NoStop}%
\bibitem [{\citenamefont {Baerlocher}\ and\ \citenamefont
  {McCusker}(2017)}]{zeo}%
  \BibitemOpen
  \bibfield  {author} {\bibinfo {author} {\bibfnamefont {C.}~\bibnamefont
  {Baerlocher}}\ and\ \bibinfo {author} {\bibfnamefont {L.}~\bibnamefont
  {McCusker}},\ }\href {\doibase http://www.iza-structure.org/databases}
  {\enquote {\bibinfo {title} {Database of zeolite structures},}\ } (\bibinfo
  {year} {2017})\BibitemShut {NoStop}%
\end{thebibliography}%

\end{document}